\documentclass[
    10pt,
    aps,
    prl,
    twocolumn,
    superscriptaddress,
    secnumarabic
]{revtex4-2}
\usepackage{braket}
\usepackage{physics}
\usepackage{qcircuit}
\usepackage{amsthm}
\usepackage{amssymb} 
\usepackage{stmaryrd} 

\usepackage{graphicx}
\graphicspath{{images/}}
\usepackage[caption=false]{subfig}

\usepackage[dvipsnames]{xcolor} 
\usepackage{hyperref}
\hypersetup{
    colorlinks=true,
    linkcolor=Maroon,
    citecolor=Maroon,
    urlcolor=Maroon,
}
\usepackage{cleveref}

\usepackage{booktabs} 

\usepackage{bm}

\usepackage{mathtools}
\usepackage{algorithm}
\usepackage{algpseudocode}

\theoremstyle{plain}
\newtheorem{theorem}{Theorem}
\newtheorem{lemma}{Lemma}

\newtheorem*{lemmastar}{Lemma 1}

\theoremstyle{definition}
\newtheorem{definition}{Definition}

\newcommand{\definitionref}[1]{%
  \hyperref[definition:#1]{Definition~\ref*{definition:#1}}%
}

\newif\ifworkmode
\workmodetrue 

\ifworkmode
  \newenvironment{observation}[1][]{%
    \par\medskip
    \noindent\textbf{\textcolor{orange}{Observation: #1}}\quad
  }{\par\medskip}
\else
  
\fi

\ifworkmode
  \newcommand{\todo}[1][]{%
    \textbf{\textcolor{red}{TO DO: #1}}%
  }
\else
  \newcommand{\todo}[1][]{}
\fi

\ifworkmode
  \newenvironment{result}[1][]{%
    \par\medskip
    \noindent\textbf{\textcolor{blue}{Result: #1}}\quad
  }{\par\medskip}
\else
  
\fi

\ifworkmode
  \newenvironment{question}[1][]{%
    \par\medskip
    \noindent\textbf{\textcolor{violet}{Question: #1}}\quad
  }{\par\medskip}
\else
  
\fi

\DeclareMathOperator{\Prob}{P}

\DeclareMathOperator*{\argmax}{arg\,max}

\DeclareMathOperator{\supp}{Supp}

\DeclareMathOperator{\rowspan}{rowspan}

\newcommand{\Hw}[1]{w_H(#1)}

\newcommand{\BB}[1]{BB#1}
\newcommand{\LLR}{\Delta}

\makeatletter
\def\equalsfill{$\m@th\mathord=\mkern-7mu
\cleaders\hbox{$\!\mathord=\!$}\hfill
\mkern-7mu\mathord=$}
\makeatother
\newcommand{\df}{=}

\newcounter{algobox}
\renewcommand{\thealgobox}{\arabic{algobox}}

\crefname{algobox}{Algorithm}{Algorithms}
\Crefname{algobox}{Algorithm}{Algorithms}

\newsavebox{\algbox}

\newenvironment{algorithmbox}
{%
  \refstepcounter{algobox}%
  \begin{lrbox}{\algbox}
  \begin{minipage}{0.95\linewidth}
  \textbf{Algorithm \thealgobox :}
}
{%
  \end{minipage}
  \end{lrbox}
  \begin{center}
    \fbox{\usebox{\algbox}}
  \end{center}
}

\definecolor{myblue}{HTML}{5677A4}
\newcommand{\highlight}[1]{\textcolor{myblue}{#1}}
\renewcommand{\thesection}{\arabic{section}}
\renewcommand{\thesubsection}{\thesection.\arabic{subsection}}
\renewcommand{\thesubsubsection}{\thesubsection.\arabic{subsubsection}}

\begin{document}
\title{Dealing with low-weight errors that lead BP OSD to fail}

\author{Giacomo Fregona}
\affiliation{Department of Computer Science, University of Copenhagen, Universitetsparken 1, 2100 Copenhagen, Denmark.}

\author{Matthias C. Löbl}
\altaffiliation{Current affiliation: Sparrow Quantum, Nordre Fasanvej 215, DK-2000 Frederiksberg, Denmark}
\affiliation{Center for Hybrid Quantum Networks (Hy-Q), Niels Bohr Institute, University of Copenhagen, Blegdamsvej 17, 2100 Copenhagen, Denmark.}

\author{Benjamin J. Brown}
\affiliation{IBM Quantum, T. J. Watson Research Center, Yorktown Heights, New York 10598, USA}
\affiliation{IBM Denmark, Sundkrogsgade 11, 2100 Copenhagen, Denmark}

\author{Michael J. Kastoryano}
\affiliation{Department of Computer Science, University of Copenhagen, Universitetsparken 1, 2100 Copenhagen, Denmark.}

\author{Stefano Paesani}
\affiliation{NNF Quantum Computing Programme, Niels Bohr Institute, University of Copenhagen, Blegdamsvej 17, 2100 Copenhagen, Denmark.}

\date{
	\today
}
 
\begin{abstract}
    \noindent 
    Belief Propagation combined with Ordered Statistics Decoding (BP OSD) is an open-source decoding tool used widely within the community due to its versatility and ease of application to Quantum Low Density Parity Check (QLDPC) codes. 
    However, the degeneracy of quantum codes has been observed to degrade the performance of BP OSD, particularly leading to error floors for low physical error rates. 
    Here, we investigate the low-weight error configurations responsible for such failures in Bivariate Bicycle codes. 
    We show that, at code capacity, BP OSD fails on errors whose weight is significantly smaller than $d/2$,  arising from global symmetries of the decoding graph that prevent the convergence of parallel BP.  
    Furthermore, our analysis reveals new pitfalls that can arise when analysing the performance of BP OSD in the rare events regime, such as loss of ergodicity in splitting methods and numerical instabilities in the decoder. 
    We show how to mitigate these failure mechanisms by modifying the BP update rules and the decoder implementation, leading to substantial improvement in logical error rate. 
\end{abstract}
\maketitle


Impactful applications of quantum computing require large numbers of noise-free logical qubits in order to perform quantum computations. 
However, reaching the physical scale required for such logical processors remains challenging for current quantum hardware. 
Quantum Low Density Parity Check (QLDPC) codes~\cite{breuckmannQuantumLowDensityParityCheck2021} enable the encoding of large numbers of logical qubits using limited physical resources, and are thus promising candidates for bridging this gap. 
Among the proposed constructions, Bivariate Bicycle (BB) codes have attracted particular attention due to their favorable finite-size performance and relatively simple structure~\cite{bravyiHighthresholdLowoverheadFaulttolerant2024}.

The performance of QLDPC codes critically depends on the availability of efficient decoders. 
BP OSD  is currently one of the most widely used decoding algorithms for QLDPC codes~\cite{panteleevDegenerateQuantumLDPC2021, roffeDecodingQuantumLowdensity2020}, and it is often regarded as the reference decoder for benchmarking new code constructions. 
BP OSD is given by the combination of two decoder, namely Belief Propagation (BP) and Ordered Statistics Decoding (OSD). 
BP relies on a network of Bayesian beliefs to update some prior error distribution based on the error syndrome. 
BP refines this distribution until convergence to a candidate error that matches the observed syndrome. 
Despite the historical success of BP in decoding classical LDPC codes~\cite{mackayShannonLimitPerformance1997}, its application to quantum codes is more problematic due to the unavoidable short cycles present in QLDPC codes~\cite{poulinIterativeDecodingSparse2008}. 
The OSD post-processor is therefore employed to tackle the degenerate cases where BP fails to converge, guaranteeing solid performance for a large class of QLDPC codes.

Despite its practical success, BP OSD inherits fundamental limitations from the underlying BP algorithm. In particular, the degeneracy of quantum codes often degrades the decoding accuracy and leads to error floors. 
Several approaches have been proposed to mitigate these issues, including modifications of the decoding graph~\cite{rigbyModifiedBeliefPropagation2019, demartiioliusAlmostlinearTimeDecoding2026}, guided or randomized message perturbations~\cite{rigbyModifiedBeliefPropagation2019, yaoBeliefPropagationDecoding2024}, alternative BP scheduling strategies~\cite{kuoRefinedBeliefPropagationDecoding2020}, memory-enhanced BP methods~\cite{chenImprovedBeliefPropagation2025}, ensemble approaches~\cite{koutsioumpasColourCodesReach2025}, and combinations thereof~\cite{mullerImprovedBeliefPropagation2025}. 
Although these strategies have been shown to improve decoding performance, they provide little insight into the fundamental mechanisms responsible for BP failures. The understanding of BP limitations therefore remains incomplete. 
This research gap is particularly remarkable at low-error rates, a regime where BP OSD has received comparatively less attention~\cite{heDiscoveringHighlyEfficient2025, landaLowweightQuantumSyndrome2026} despite being of greatest practical relevance~\cite{kimFaulttolerantResourceEstimate2022, gidneyHowFactor20482025}.  
A deeper understanding of these failures can enable us to develop better decoding algorithms that overcome these issues, interpret the decoder's parameter space, and compare existing approaches.

\begin{figure*}
    \centering
    \includegraphics[width=1\linewidth]{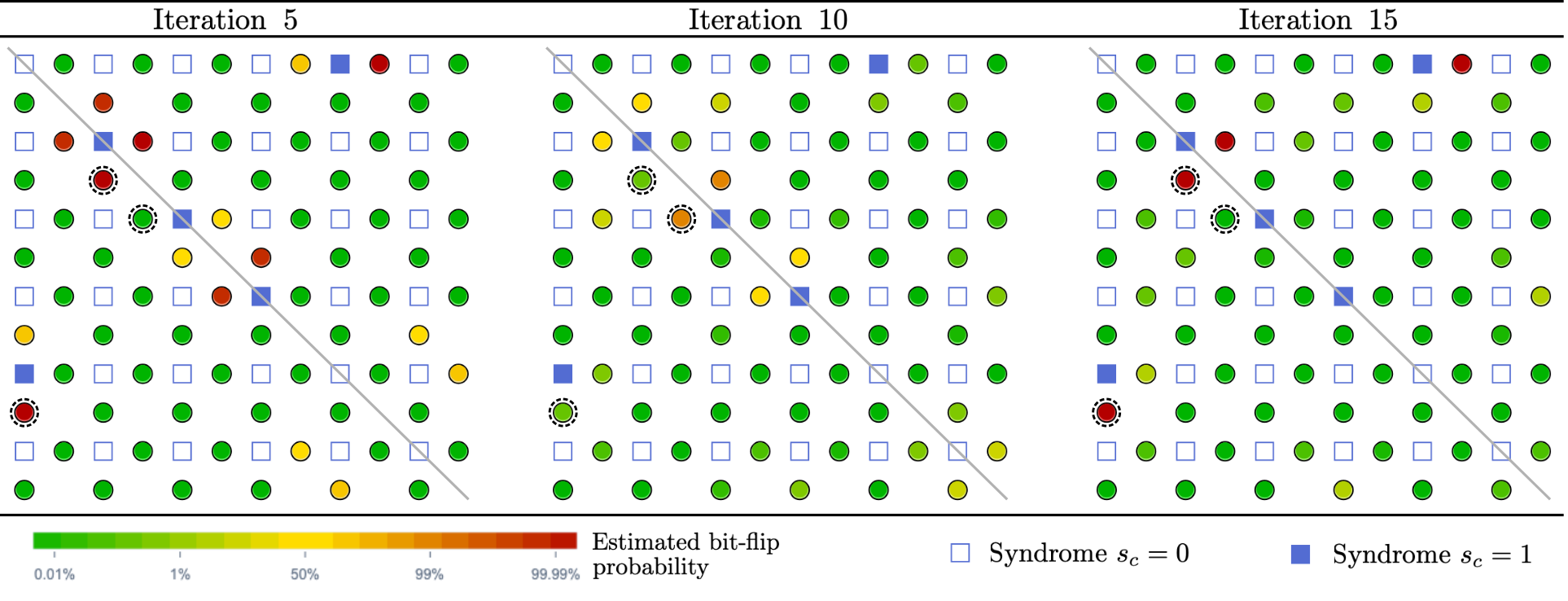}
    \caption{
        Symmetric soft-information induced by reflections, BB code with parameters $[\![72,12,6]\!]$ from ref.~\cite{bravyiHighthresholdLowoverheadFaulttolerant2024}. The image represents the soft information produced by parallel BP at iterations $t = 5, \;10, \; 15$ in decoding a chosen weigh-$3$ error (dashed circles). Qubit (dots) and checks (squares) representation and ordering is analogous to fig. 1 from ref.~\cite{bravyiHighthresholdLowoverheadFaulttolerant2024}. BP's soft decision is represented by the qubit colors. The soft decision is symmetric with respect to a reflection along the diagonal axis.
    }
    \label{fig:BB72_iterations}
\end{figure*}

\textbf{Contributions.}
In this work, we demonstrate that BP OSD fails on error patterns of unexpectedly low weight which induce symmetries in BP's soft-information (see \cref{fig:BB72_iterations}). Furthermore, we show that these failures can be avoided by modifying BP's update rule without compromising the efficiency and the overall performance of the decoder.
To identify these errors, we combined exhaustive numerical simulations and theoretical analysis, characterizing low-weight failures of BP OSD for BB codes at code capacity. In these investigations, we found two obstacles that complicate the assessment of BP OSD's performance, given by numerical instability in the decoder implementation~\cite{roffeDecodingQuantumLowdensity2020} and lost ergodicity in the Markov chain underlying the splitting method~\cite{bravyiSimulationRareEvents2013, mayerRareEventSimulation2025, beverlandFailFastTechniques2025}.
Overall, our work provides a deeper understanding of the mechanism causing failure of BP OSD, and underlines potential pitfalls in analyzing BP OSD at very low physical error rate.

\section{Notation and Background}
This section introduces the notation and definitions used throughout the paper, with particular emphasis on the intuition underlying the BP OSD decoder. We first introduce the standard definitions of Tanner graphs and quantum CSS codes in \cref{subsec:def}, as well as the notation used for Bivariate Bicycle codes. For a more complete treatment of these concepts, we refer the reader to appendix \ref{sec:definitions}. 
\Cref{subsec:decoding_framework}  discusses the decoding problem in general terms and then specializes it to the bit-flip noise model considered in our simulations. 
\Cref{subsec:BP} presents the BP decoder. Here, our exposition of BP centers on the progressive approximations of the decoding problem that motivate the decoder's update rules, aiming to precisely specify why BP decoding is heuristic.
Finally, although our findings primarily concern BP, we include a brief description of the standard OSD post-processing used in our simulations in \Cref{subsec:OSD}.

\subsection{Preliminary Definitions}\label{subsec:def}

We begin with basic notions of classical coding theory, which  provide the foundation for the definitions of Tanner graphs and quantum CSS codes.

The ambient space for our discussion is the binary vector space $\mathbb{F}_2^n$. 
Following the convention used in coding theory, vectors $x\in\mathbb{F}_2^n$ are regarded as row vectors. We use $\Hw{x}$ to denote the (Hamming) weight of $x\in \mathbb{F}_2^n$, and the symbol $\oplus$ for addition modulo $2$. The symbol $\supp(x)$ denotes the support of $x\in\mathbb{F}_2^n$, that is, the set of indexes corresponding to nonzero coordinates of $x$.

A classical (binary) $[n,k,d]$-code $\mathcal{C}$ is a $k$-dimensional subspace of a binary vector space $\mathbb{F}_2^n$. The parameter $d$, called distance, is the minimum weight of nonzero vectors in $\mathcal{C}$. 
A parity check matrix $H\in \mathbb{F}_2^{m\times n}$ is a matrix whose rows generate the orthogonal complement of the code, for which we use the notation $\mathcal{C}^\perp$. By construction, $m \geq n - k$. Membership to a code $x\in \mathcal{C}$ can be tested using the equivalent identity $x H^t = 0$.

In our discussion, it will be often convenient to describe the parity check matrix of a classical code with the alternative representation of Tanner graph $\mathcal{T}=(V,C,E)$. The graph $\mathcal{T}$ is bipartite, and it has two sets of nodes: $V$, whose elements are called variables, and $C$, whose elements are called checks. The undirected edges $E \subseteq V \times C$ are determined by the parity check matrix of the graph $H\in \mathbb{F}_2^{m\times n}$. More precisely, $H$ is the adjacency matrix of the graph. In this context, it is convenient to identify $\mathbb{F}_2^n$ with $\mathbb{F}_2^V$, where $\mathbb{F}_2^V$ is the set of vectors whose coordinates are labelled by the vertices $V$. Similarly, we identify $\mathbb{F}_2^m$ with $\mathbb{F}_2^C$. The symbol $N(w)$ denotes the set of vertices adjacent to $w\in V\cup C$.

We now introduce the notion of a quantum CSS code. A $[\![n,k,d]\!]$-quantum CSS code is a pair of classical codes $\mathcal{C}_X,\mathcal{C}_Z \subseteq \mathbb{F}_2^n$ such that $\mathcal{C}_X^\perp \subseteq \mathcal{C}_Z$. Despite the fact that quantum codes are described in terms of binary codes, the binary vectors used in their description do not merely represent bit strings, but also provide a representation of quantum operators belonging to the $n$-qubit Pauli group $\mathcal{P}_n$. In particular, up to an overall phase, any Pauli operator can be represented by a binary vector $(x\mid z)\in\mathbb{F}_2^{2n}$, corresponding to
\[
P(x\mid z)=X^xZ^z
=\prod_{i=1}^n X_i^{x_i}Z_i^{z_i}.
\]
For CSS codes, the relevant operators are of either $X$-type or $Z$-type, and can therefore be represented by vectors in $\mathbb{F}_2^n$. \\
CSS codes inherit their terminology from their physical interpretation. The parameters $n$ and $k$ are respectively called the number of physical qubits and the number of logical qubits of the code. The number of logical qubits is given by
$
k = n-\dim(\mathcal{C}_X^\perp)-\dim(\mathcal{C}_Z^\perp).
$
The relation $\mathcal{C}_X^\perp\subseteq\mathcal{C}_Z$ ensures the commutation of the measurement operators used to protect the quantum information. These operators, corresponding to vectors in $\mathcal{C}_X^\perp$ and $\mathcal{C}_Z^\perp$, are called $X$- and $Z$-stabilizers, respectively. Elements in $\mathcal{C}_Z\setminus\mathcal{C}_X^\perp$ and $\mathcal{C}_X\setminus\mathcal{C}_Z^\perp$ instead represent logical operations on the quantum information protected by the code. They are called logical-$X$ and logical-$Z$ operators, respectively. The distance $d$ of a CSS code is the minimum weight of a vector in $(\mathcal{C}_Z \setminus \mathcal{C}_X^\perp) \cup (\mathcal{C}_X \setminus \mathcal{C}_Z^\perp)$, that is, the minimum weight of a logical operator.

We will reserve $H_X, H_Z \in \mathbb{F}_2^{m \times n}$ to denote the parity check matrices of $\mathcal{C_X}$ and $\mathcal{C}_Z$. The logical action on quantum CSS codes can be described using two logical action matrices $A_X, A_Z  \in\mathbb{F}_2^{k \times n}$. The two matrices $A_X$ and $A_Z$ satisfy the two conditions
\begin{align*}
    \mathcal{C}_Z & = \rowspan(H_X) \oplus \rowspan(A_X), \\
    \mathcal{C}_X & = \rowspan(H_Z) \oplus \rowspan(A_Z),
\end{align*}
where $\oplus$ denotes the direct sum of linear subspaces.
In the context of a quantum CSS code, we will use the notation $\mathcal{T}_X=(Q,C_X,E_X)$, $\mathcal{T}_Z=(Q,C_Z,E_Z)$ for the Tanner graphs associated to the parity check matrices $H_X$ and $H_Z$, where the variable set $Q$ is common to both the Tanner graphs and is called set of qubits. We will refer to $C_X$ and $C_Z$ as the set of $X$-checks and $Z$-checks respectively.

We considered a special class of quantum CSS codes, namely Bivariate Bicycle codes \cite{bravyiHighthresholdLowoverheadFaulttolerant2024}. We refer the reader to appendix \ref{sec:definitions} for a definition of BB codes. In \cref{table:reflections} we use the notation for qubits and checks of BB codes introduced in \cite{bravyiHighthresholdLowoverheadFaulttolerant2024}. We briefly recall this notation below.\\
Every BB code is associated with two natural numbers $l,m \in \mathbb{N}$. Let $\mathcal{M}$ be the following set of bivariate monomials
\[
    \mathcal{M} = \{x^a y^b \mid a < l, b < m\} \subseteq\mathbb{F}_2[x,y].
\]
We also use the symbols $L, R, X, Y$, which here serve as labels. Qubits in BB codes are conventionally identified by couples of the form $(L, x^a y^b)$ or $(R, x^a y^b)$, where $x^a y^b \in \mathcal{M}$. Similarly, $X$- and $Z$-checks are labelled by $(X, x^a y^b)$ and $(Z, x^a y^b)$, respectively, with $x^a y^b \in \mathcal{M}$. 

\subsection{Decoding Problem}\label{subsec:decoding_framework}
Central to the decoding framework is the definition of the decoding problem. We introduce here a general formulation of this problem \ref{def:decoding_problem}, closely following the approach of \cite{ottDecisiontreeDecodersGeneral2025}. In \cref{bit_flip}, we specialize this formulation to the bit-flip noise model considered in our simulations. We additionally introduce the standard definition of the logical error rate and the optimal decoding strategy obtained by minimizing it, as well as Most Likely Error (MLE) decoding, which provides an approximation to this optimal strategy and will be central to our discussion of BP.
\begin{definition}[Decoding problem]\label{def:decoding_problem}
    Let $H \in \mathbb{F}_2^{m \times n}$ be a \emph{parity check matrix} and $A\in \mathbb{F}_2^{l \times n}$ be a \emph{logical action matrix}. Let the errors $x \in \Omega =\mathbb{F}_2^n$ be sampled according to some probability $P$. Considering an unknown error $x\in\mathbb{F}_2^n$, and given a syndrome $s = x H^T$, the decoding problem consists in providing a correction $\tilde{x}\in\mathbb{F}_2^n$ such that $\tilde{x}H^T = s$ and $\tilde{x}A^T = xA^T$. 
\end{definition}

Classical algorithms designed to solve the decoding problem are called \emph{decoders}. For an $x\in \Omega$, if a decoder runs with input syndrome $x H^T$ and outputs a correction that does not satisfy the conditions listed in the definition, we say that $x$ induces a failure of the decoding procedure. For a particular decoder, we denote with $\mathcal{F} \subseteq \Omega$ the set of errors inducing the failure of the decoding.

Throughout this work, we will always assume that $\Prob$ takes the particular form of 
\begin{equation}\label{equation:probability_distribution}
    \Prob(x) = p^{\Hw{x}} (1-p)^{n-\Hw{x}}\text{ for every } x\in \Omega,
\end{equation}
or equivalently, the entries of $x$ are drawn according to independent identically distributed Bernoulli random variables with parameter $p\in [0,1]$. The value $p$ is called \emph{physical error rate} of the decoding problem. 

A metric of performance for a decoder is the \emph{Logical Error Rate} (LER), which is defined as the probability measure $\Prob(\mathcal{F}) = \sum_{x \in \mathcal{F}} \Prob(x)$. \\
There exist an optimal strategy to minimize the LER. For every syndrome $s\in\mathbb{F}_2^m$ and $y\in\mathbb{F}_2^l$, define
\[
    C_y^s = \{x\in\mathbb{F}_2^n \mid xH^T = s, \; xA^T = y\}.
\]
Consider a decoder which selects $\tilde{x}\in C_{y'}^s$, with $y' = \argmax_{y} \left[\sum_{x \in C_y^s}P(x)\right]$. This decoder is optimal, and known as \emph{Most Likely Coset} decoder (MLC).
Despite the optimality of MLC decoders, they do not admit feasible implementation in general. For this reason, as already argued in \cite{dennisTopologicalQuantumMemory2002}, an approximate solution can be obtained maximizing the likelihood of an individual error $x\in \Omega$. For a syndrome $s$, the likelihood of $x$ is given by
\[
    \Prob(x \mid s) = 
    \begin{cases}
        \frac{\Prob(x)}{Z} & \text{ if } xH^t = s \\
        0 & \text{otherwise,}
    \end{cases}
\]
where $Z$ is a normalization factor. Decoders that select $\tilde{x} = \argmax_{x \in \Omega} \Prob(x \mid s)$ are called \emph{Most Likely Error} decoders (MLE). With the particular choice \ref{equation:probability_distribution} of the probability function, maximizing $P(x \mid s)$ is equivalent to minimizing the weight of the error $x$ within the class of errors satisfying the syndrome. We call this strategy \emph{minimum weight decoding}.

In our simulations, the considered the following specialized version decoding problem, obtained from CSS codes.
\begin{definition}[Bit-flip decoding problem] \label{bit_flip}
    Let $H_X, H_Z \in \mathbb{F}_2^{m \times n}$ be the parity check matrices of a quantum CSS code, and let $A_Z$ be a logical-$Z$ action matrix. The bit-flip decoding problem is a particular case of the decoding problem where $H \df H_Z$ and $A = A_Z$.
\end{definition}

For an error $x \in \mathbb{F}_2^n$ and a candidate correction $\tilde{x}$, we can rephrase the conditions for successfull decoding in terms of their composition $c = x \oplus \tilde{x}$ as
\[
    c H^t_Z = 0 \text{ and } c A_Z^t = 0.
\]
By definition of the parity check matrix $H_Z$, the requirement $c H^t_Z = 0$ holds precisely when $c \in \mathcal{C}_Z$. By construction of $A_Z$, the two conditions altogether are equivalent to $c \in \mathcal{C}^\perp _X$. \\
We can therefore characterize the success of an instance of the bitflip decoding problem depending on $c$. Decoding fails if $c \notin \mathcal{C}_Z$, that is, if the correction's syndrome is different from the input syndrome. Decoding alternatively fails if $c \in \mathcal{C}_Z \setminus \mathcal{C}_X ^\perp$, meaning that $c$ is an $X$-logical operator of the code. Finally, decoding succeeds when $c \in \mathcal{C}_X ^\perp$, namely $c$ is an $X$-stabilizer of the code.

Considering a minimum weight decoder, the distance $d$ of the CSS code imposes a bound on the weight of the errors that induce failuire of the decoding procedure. Let $x$ be the original error and $\tilde{x}$ the solution proposed by a minimum weight decoder. By definition of minimum weight decoder, the proposed solution matches the input syndrome. In the bit-flip decoding problem, this implies that $c = x \oplus \tilde{x}$ is either an $X$ stabilizer or an $X$-logical operator of the code. Additionally, minimum weight decoding guarantees that $\Hw{\tilde{x}} \leq \Hw{x}$, and therefore the weight of $c$ is limited by 
\[
    \Hw{c} = \Hw{x \oplus \tilde{x}} \leq \Hw{x} + \Hw{\tilde{x}} \leq 2\Hw{x}
\]
If the decoding procedure fails, necessarily $c$ is an $X$-logical operator of the code. The distance $d$ is a lower bound on the weight of logical operators, and therefore $d \leq \Hw{c} \leq 2 \Hw{x}$, or equivalently, $\Hw{x} \geq \left\lceil\frac{d}{2}\right\rceil$. In other terms, minimum weight decoding provides the correct solution for every error whose weight is less than or equal 
\begin{equation}\label{weight_limit}
     \left\lceil\frac{d}{2}\right\rceil - 1 = \left\lfloor \frac{d-1}{2} \right\rfloor.
\end{equation}
Since BP OSD is an approximation of MLE decoding, which in turn is equivalent to minimum weight decoding in the studied cases, we use the weight limit in \cref{weight_limit} as a reference for evaluating the performance of BP OSD in our numerical experiments.

\subsection{BP Decoding}\label{subsec:BP}
Let us now introduce the BP decoder. From a broad perspective, Belief Propagation assigns to each variable an estimated error probability, referred to as the soft decision, and uses this estimate to produce candidate solutions. In this section, we first specify which probability is  approximated by the soft decision (\cref{equation:marginal}). We subsequently describe how BP makes use of this information to propose a solution to the decoding problem. Next, \cref{theorem:gallagher} establishes the relation between the marginal probability in \cref{equation:marginal} and the estimated soft decision, clarifying the approximations that lead to the formulation of BP. The algorithmic computation of the soft information is described separately in box \ref{box:BP} at the end of the section.


In order to introduce BP, consider a decoding problem, and let $p$ be the associated physical error rate. Let $\mathcal{T} = (V,C,E)$ be the Tanner graph associated to the parity check matrix $H$. Let  $x\in \Omega$ and $s = xH^T$. \\
The value
\begin{equation}\label{equation:marginal}
    \pi_v \df \sum_{x\in\Omega \mid x_v = 1} P(x \mid s)
\end{equation}
represents the marginal of the likelihood function for the variable $v\in V$. The strategy underlying BP is to approximate these marginals $\tilde{\pi}_v \sim \pi_v$ efficiently and produce a candidate solution $\tilde{x} = (\tilde{x}_v)_{v\in V}$ with
\begin{equation}\label{hard_decision}
    \tilde{x}_v \df 
    \begin{cases}
        1& \text{if } \tilde{\pi}_v > \frac{1}{2},\\
        0& \text{otherwise.}
    \end{cases}
\end{equation}
BP is therefore an approximation of MLE decoding. The estimate $\tilde{\pi}_v$ is called \emph{soft information}, or \emph{soft decision}, while a candidate $\tilde{x}$ chosen according to \cref{hard_decision} is commonly known as \emph{hard decision}. The BP algorithm computes and iteratively refines the soft and hard decision. In relation to an iteration counter $t \in \mathbb{N}$, we can therefore talk about hard and soft decision at iteration $t$.

The procedure for the computation of $\tilde{\pi}_v$ is defined in \cref{box:BP}. The following theorem, which is a restatement of the classical result due to Gallagher \cite{LowDensityParityCheckCodes1963}, specifies more precisely the relation between $\pi_v$ and the soft decision $\tilde{\pi}_v$. 

\begin{theorem}\label{theorem:gallagher}
    Consider a decoding problem with associated Tanner graph $\mathcal{T} = (V,C,E)$. Let $\tilde{\pi}_v$ be the soft decision for the variable $v$, produced by the BP algorithm \ref{box:BP} at iteration $t \in \mathbb{N}$ with input syndrome $s$. Let $\mathcal{T}' = (V',C',E')$ be the sub-Tanner graph of $\mathcal{T}$ having nodes with distance 
    \footnote{
    Here, the distance function $d$ is the minimum number of edges in a path connecting two vertices of the Tanner graph (graph distance). In greater detail, $V' = \{v' \in V \mid d(v, v') \leq 2t\}$, $C' = \{c' \in C \mid d(v, c') \leq 2t\}$ and $E'= E \cap (V' \times C')$.
    } from $v$ bounded by $2t$, and $s' \df (s_c)_{c\in C'}$. If $\mathcal{T}'$ is acyclic, then\begin{equation}\label{equation:soft_information}
        \tilde{\pi}_v = \sum_{x'\in\mathbb{F}_2^{V'} \mid x'_v = 1} P(x' \mid s').
    \end{equation}
\end{theorem}
The right-hand side of \cref{equation:soft_information} represents the marginal of the form \ref{equation:marginal}, computed using a local restriction of the information brought by $s$. The computation at successive iteration of the BP decoder involves therefore additional information from the syndrome in the estimate of the marginals. Hence the estimate incorporates the full information carried from the syndrome if $t$ is large enough that $\mathcal{T} = \mathcal{T}'$. Despite that, the common practice in BP decoding is to allow much larger values of $t$, following the intuition that a progressively refined estimates of the marginals $\pi_v$ lead to an improved hard decision guess.

Belief propagation is not derived directly from the optimal decoding problem, but rather through a sequence of simplifying assumptions and algorithmic approximations. Since our work focuses on the limitations and failure mechanisms of BP, we conclude this section by making these approximations explicit. 
\begin{enumerate}
    \item BP follows the Most Likely Error decoding approach, which is in general different from optimal decoding.
    \item \label{marginalization}
    The marginalization strategy selecting $\tilde{x}$ does not guarantee that the candidate solution maximizes $P(\tilde{x}|s)$ globally.
    \item \label{acyclicity}
    The computation of the marginals is based on an acyclicity assumption, which is rarely met in the cases of application of BP.
\end{enumerate}
As a consequence, the use of BP is considered a heuristic. In the decoding of quantum CSS codes, the performance of BP in terms of logical error rate and minimum weight error is often significantly worse compared to the potential error correction capability of Most Likely Error decoding~\cite{poulinIterativeDecodingSparse2008, rigbyModifiedBeliefPropagation2019, raveendranTrappingSetsQuantum2021}. For this reason, BP is often coupled with a post-processing algorithm. The OSD decoder, described in the next section, is the standard choice for BP post-processing. 

\begin{algorithmbox}\label{box:BP}
\textbf{Belief Propagation Decoding}\\\\
BP is based on a network of messages which are iteratively exchanged along the edges of the Tanner graph $\mathcal{T} = (V,C,E)$. \\
Let $t\in\mathbb{N}$ denote the iteration counter of the algorithm. For every $t\in\mathbb{N}$ and $(c, v) \in E$, the algorithm exchanges two messages, called \emph{beliefs}. These beliefs are known as \emph{variable-to-check} ($m^t_{v\to c}$) and \emph{check-to-variable} ($m^t_{c\to v}$) messages. They represent a probability $\pi\in[0,1]$ encoded $\pi$ in the form of \emph{Log Likelihood Ratio} $\LLR(\pi) = \ln(\frac{1-\pi}{\pi})$.

At $t=0$, $m^0_{v\to c} = \LLR(p)$, where $p$ is the considered physical error rate. Check-to-bit messages $m^t_{c\to v}$ are not defined for $t=0$. \\
For $t \geq 1$, the algorithm performs the following steps.
\begin{enumerate}
    \item \label{BP_step_1}
    For every edge $(c,v) \in E$, compute the messages:
        \begin{align}\label{BP_update_rule}
            m^t_{c\to v} &= (-1)^{s_c} f^{-1}\left[\prod_{v'\in N(c)\setminus \{v\}} f(m^{t-1}_{v' \to c})\right],\\
            m^t_{v \to c} &= \LLR(p) + \sum_{c'\in N(v)\setminus \{c\}} m^t_{c' \to v},
        \end{align}
    with $f(\pi) = \tanh(\pi/2)$.

    \item \label{BP_step_2}
    For every variable $v \in V$, compute the estimate $\tilde{\pi}_v = \LLR^{-1}(d^t_v)$, where:
    \begin{equation}\label{BP_soft_decision}
        d^t_v = \LLR(p) + \sum_{c\in N(v)} m^t_{c \to v}.
    \end{equation}
    \item  \label{BP_step_3}
    Select $\tilde{x}$ based on the procedure \ref{hard_decision}. If $\tilde{x} H^t = s$, halt and return $\tilde{x}$.
\end{enumerate}
If a maximum iteration counter $T$ is reached, the decoder terminates outputting a failure status.
\end{algorithmbox}

\subsection{OSD Decoding}\label{subsec:OSD}
Ordered Statistics Decoding (OSD) is the decoding algorithm commonly used as a post-processing step for BP, together forming the BP OSD decoder. OSD was originally introduced as an independent decoder for classical codes~\cite{fossorierReliabilitybasedSyndromeDecoding1998}, and later adapted to quantum CSS codes~\cite{panteleevDegenerateQuantumLDPC2021}. In the remainder of this section, we provide a brief description of the algorithm following~\cite{roffeDecodingQuantumLowdensity2020}.

Consider the decoding problem from \cref{def:decoding_problem}, and let $V$ denote the set of variable labels. Given an input syndrome $s$, suppose that, for every $v\in V$, the OSD decoder receives an estimate $\tilde{\pi}_v$ of the marginal probability $\pi_v$. Let $\tilde{x}$ denote the corresponding hard decision obtained as in \cref{hard_decision}. For every $v\in V$, the algorithm computes the reliability score $r_v = 1-\tilde{\pi}_v$
\footnote{
The two versions of OSD described in~\cite{panteleevDegenerateQuantumLDPC2021} and~\cite{roffeDecodingQuantumLowdensity2020} differ in the definition of the reliability score. In particular,~\cite{panteleevDegenerateQuantumLDPC2021} uses $r_v=\max(\tilde{\pi}_v,1-\tilde{\pi}_v)$, whereas our simulations follow the definition adopted in~\cite{roffeDecodingQuantumLowdensity2020}.
}
which quantifies the confidence in the hard decision.

The variables are then sorted in decreasing order of reliability, and Gaussian elimination is applied to the parity-check matrix $H$ according to this ordering. This identifies an information set $I\subseteq V$~\cite{prangeUseInformationSets1962}, corresponding to the set of pivot columns obtained during Gaussian elimination. The values assigned to the variables in $I$ uniquely determine the remaining variables through the syndrome equation $xH^t=s$. Therefore, the restriction $\tilde{x}_I=(\tilde{x}_v)_{v\in I}$ of the hard decision to $I$ is used to obtain a candidate solution $x'$ such that $x'_I=\tilde{x}_I$ and $x'H^t=s$.

The decoder can be strengthened by introducing an additional parameter $q\in\mathbb{N}$, called order. Order-$q$ OSD generates additional candidate solutions by exhaustively flipping up to $q$ entries of $\tilde{x}_I$, re-encoding each modified information-set assignment, and selecting the candidate of minimum Hamming weight. Unless otherwise stated, all simulations in this work use order-$0$ OSD for computational efficiency. Nevertheless, in the rare-events analysis we discuss how the identified failure mechanisms extend to higher OSD orders.
\section{Numerical Simulations}
Here we describe our numerical experiments. Specifically, \cref{subsec:methods} outlines the simulation methodology and explains how the simulation outcomes are used to estimate the logical error rate of the decoder, together with a mathematical bound on the estimation error. We refer the reader to appendix \ref{error_bound} for a detailed derivation of this bound.

\Cref{subsec:parameters} specifies the Bivariate Bicycle (BB) codes considered in our simulations. For further clarifications on BB codes and their defining parameters, we direct the reader to appendix \ref{sec:definitions}. The same section also details the decoder implementation and the BP input parameters used throughout our simulations.

Conducting our simulations, we faced an issue related to  numerical instability in the decoder implementation. \Cref{numerical_instability} describes this complication, that we believe could be relevant for other works within the scope of QEC. We also detail the methodology used to circumvent this issue and eliminate its impact from our results.  

Finally, the simulation results are presented and discussed in \cref{simulations_results}. The set of failure errors $\mathcal{F}$ for codes \BB{144} and \BB{288} exhibit peculiar configurations which would be problematic to analyze using Metropolis-based techniques such as the Splitting method~\cite{bravyiSimulationRareEvents2013, mayerRareEventSimulation2025, beverlandFailFastTechniques2025}. This is argued in the second paragraph of the section.

\subsection{Methods}\label{subsec:methods}

In our simulations, we exhaustively tested the decoding failures of low weight errors. More precisely, let $\Omega_w \subset \mathbb{F}_2^n$ denote the set of errors of weight $w$ for a given decoding set. We decoded the syndrome $s = x H^T$ for every error $x \in \Omega_w$ and verified the success or failure of the decoding procedure for $w \leq 5$ \footnote{The case $w = 5$ for the code \BB{288} is excluded to maintain computational tractability.}. Let $N_w$ denote the number of decoding failures of weight $w$, that is, $N_w = |\mathcal{F} \cap \Omega_w|$. Our simulations provide us the exact count $N_w$ for the considered weights $w$.

We used the obtained counts to determine exactly the minimum weight of an error inducing decoding failure, and to estimate the logical error rate $\Prob(\mathcal{F})$.\\
To approximate $\Prob(\mathcal{F})$, we considered the expression
\begin{multline}\label{eq:LER_estimate}
    \Prob(\mathcal{F}) = \sum_{w = 1} ^ n N_w \;p^w (1-p)^{n-w} = \\
    = \left[\sum_{w = 1} ^ 5 N_w \;p^w (1-p)^{n-w}\right] + r,
\end{multline}
where $r = \sum_{w = 6} ^ n N_w \;p^w (1-p)^{n-w}$ captures the higher-weights contribution. We computed the first term exactly and derived a rigorous bound on $r$:
\begin{equation} \label{eq:bound}
    r \leq \beta^{6} \frac{1-\beta^{n-5}}{1-\beta},
\end{equation}
where $\beta = \frac{n e p }{6}$. The derivation of the bound is described in \cref{error_bound}. \\

Our estimation method can be seen as an application of importance sampling~\cite{smithQuickSimulationReview1997, beverlandFailFastTechniques2025}, where the values $N_w$ are exactly determined.
Our simulations are based on fixed-weight sampling, in contrast to standard Monte Carlo algorithms which fix the physical error probability instead. This choice improves computational efficiency, similarly to the use of the microcanonical ensemble in related problems in statistical physics~\cite{newmanFastMonteCarlo2001}.

\subsection{Codes and Parameters}\label{subsec:parameters}
\textbf{BB codes.}  We considered the bit-flip decoding problem obtained from five different Bivariate Bicycle codes from~\cite{bravyiHighthresholdLowoverheadFaulttolerant2024}, whose parameters are reported in \cref{table:BB_codes_parameters}. For convenience, we will name these codes by their number of physical qubits as \BB{72}, \BB{90}, \BB{108}, \BB{144} and \BB{288}.

\textbf{Noise model.} In contrast to ciruit level noise simulations, the bit-flip noise model makes the simplification of assuming that syndrome measurements are perfect. This keeps simulations feasible, despite their exhaustive nature, and represents the simplest setup for the discussion of the error mechanisms identified in our analysis.

We selected the physical error rate $p = 10^{-4}$. This choice collocates us below the pseudo-threshold of all the considered code instances. This configuration is necessary to keep the contribution of high-weight errors small ($r$ in \cref{eq:LER_estimate}), and therefore justify the precision in our estimate of the logical error rate. We remark that this choice of $p$ keeps our investigations in the sub-threshold regime that is particularly relevant for future applications of quantum computing~\cite{kimFaulttolerantResourceEstimate2022, gidneyHowFactor20482025} and less studied due to the ineffectiveness of Monte Carlo based simulations for low logical error rate.

\textbf{Decoder parameters.} Our work builds upon the BP OSD decoder implemented in~\cite{roffeDecodingQuantumLowdensity2020}. We used parallel, product-sum BP, with maximum $T = 500$ iterations. We set the input physical error rate of the decoder to the same value $p = 10^{-4}$ used for sampling. We applied order $0$ OSD post-processing.

\begin{table}[ht]
    \centering
    \begin{tabular}{l|c|c|c|c|c}
        \hline
        Code & $l$ & $m$ & $A$ & $B$  &  Code Parameters\\
        \hline
        \BB{72} & $6$ & $6$ & $x^3+y+y^2$ & $y^3+x+x^2$  &  $[\![72,12,6]\!]$\\

        \BB{90} & $15$ & $3$ & $x^9+y+y^2$ & $1+x^2+x^7$  &  $[\![90,8,10]\!]$ \\

        \BB{108} & $9$ & $6$ & $x^3+y+y^2$ & $y^3+x+x^2$ & $[\![108,8,10]\!]$  \\

        \BB{144} & $12$ & $6$ & $x^3+y+y^2$ & $y^3+x+x^2$ & $[\![144,12,12]\!]$ \\

        \BB{288} & $12$ & $12$ & $x^3+y^2+y^7$ & $y^3+x+x^2$ & $[\![288,12,18]\!]$  \\
        \hline
    \end{tabular}
    \caption[Parameters of the Bivariate Bicycle codes considered in our simulations.]
    {Parameters of the Bivariate Bicycle codes considered in our simulations, taken from~\cite{bravyiHighthresholdLowoverheadFaulttolerant2024}.}
    \label{table:BB_codes_parameters}
\end{table}

\subsection{Numerical Instability in BP OSD}\label{numerical_instability}
Our numerical study revealed a numerical stability issue in the BP OSD implementation we used~\cite{roffeDecodingQuantumLowdensity2020}. Because this implementation is widely used in the QEC community, the issue may also be relevant for other numerical studies. We therefore describe the problem and explain how we ruled out its impact on our results.

By inspecting the network of BP's beliefs, we observed that check-to-variable messages may attain infinite values $m^t_{c\to v} =\pm \infty$. Therefore, if the variable-to-check rule \ref{BP_update_rule} combines two infinite check-to-variable beliefs of opposite sign, the resulting message results in an arithmetic error (Not a Number).
In this eventuality, the numerical issue spreads along the messages of the decoding graph, corrupting the functioning of the decoder, which runs until the maximum number of iterations and produces useless soft-information. As a consequence, OSD's success is a matter of chance, determined by the collocation of the considered error in the default ordering of variables.

In order to exclude the effect of the numerical stability from our results, we introduced a minimal variation in the BP algorithm. After the computation of every variable-to-check message, we test whether it represents an arithmetic error, and possibly replace the error with the value $\LLR(\frac{1}{2}) =0$. We chose this value because it represents the maximally uncertain estimate ($50\%$ error probability). This variation does not eliminate the cause of the numerical instability, but prevents the arithmetic errors from spreading. Despite more complete solutions of the numerical issue are possible, we chose this minimal variation to demonstrate simply and directly the impact of the numerical instability. We remark that ceiling the check-to-variable beliefs would induce a similar dynamic. \\
The results presented in \cref{table:min_weight_simulations} are obtained using this variation of the BP algorithm. The potential impact of numerical instability is instead represented in \cref{figure:counters_weight_34}. Here, its contribution is measured by taking the difference between the error counts of the numerically unstable version and the modified algorithm. The same contribution, measured in terms of logical error rate, is shown in \cref{fig:decoder_performance}.

\subsection{Simulation Results}\label{simulations_results}
The results of our simulations, consisting in the counts $N_w$, are reported in \cref{table:min_weight_simulations}. Remarkably, the simulated minimum weight of failure inducing errors is $w=3$ for codes \BB{144} and \BB{288}. The comparison with the expected minimum weight $\lfloor \frac{d-1}{2}\rfloor + 1$ shows an evident mismatch between the actual decoder performance and the potential error correction capability of these codes. This difference highlights a limitation of the decoding algorithm.
\begin{table}[ht]
    \centering
    \begin{tabular}{l|c|c|c|c|c}
        \hline
        Code  &   $N_1$ & $N_2$ &$N_3$ & $N_4$ & $N_5$\\
        \hline
        \BB{72}  &0&0& 1.31e3 & 7.80e4 & 3.38e6\\

        \BB{90} &0&0& 0 & \highlight{5.33e2} & 1.91e5\\

        \BB{108} &0&0& 0 & \highlight{4.88e2} & 7.30e4\\

        \BB{144}  &0&0& \highlight{1.26e2} & 0 & \highlight{2.67e3}\\

        \BB{288} &0&0& \highlight{8.64e2} & 0 & -\\
        \hline
    \end{tabular}
    \caption{Counts $N_w$ obtained from our simulations. $N_w$ is the number of weight-$w$ errors that induce a decoding failure. Values that are incompatible with the expected minimum weight $\lfloor \frac{d-1}{2}\rfloor + 1$ are colored in blue. These results are obtained from the numerically stable variation of BP OSD.}
    \label{table:min_weight_simulations}
\end{table}

Again for codes \BB{144} and \BB{288}, the decoder fails on some weight-$3$ errors, but always retrieves the proper solution for every weight-$4$ configuration. We observe that this specific scenario is impossible to analyze using the Splitting Method~\cite{bravyiSimulationRareEvents2013, mayerRareEventSimulation2025, beverlandFailFastTechniques2025}. This technique applies a Metropolis procedure to explore the space of failures $\mathcal{F}$. Each Metropolis step transitions from errors of weight $w$ to errors of weight $w\pm1$. Due to the layer of fully correctable errors of weight $4$, the class of weight-$3$ errors is disconnected from the higher weight failures in $\mathcal{F}$. This configuration violates the ergodicity assumption required by the Splitting. When applied to this system, the method would neglect the contribution of weight-$3$ failures, and underestimate the logical error rate of the decoder by orders of magnitude.

\begin{figure}[h]
    \centering
    \includegraphics[width=1\linewidth]{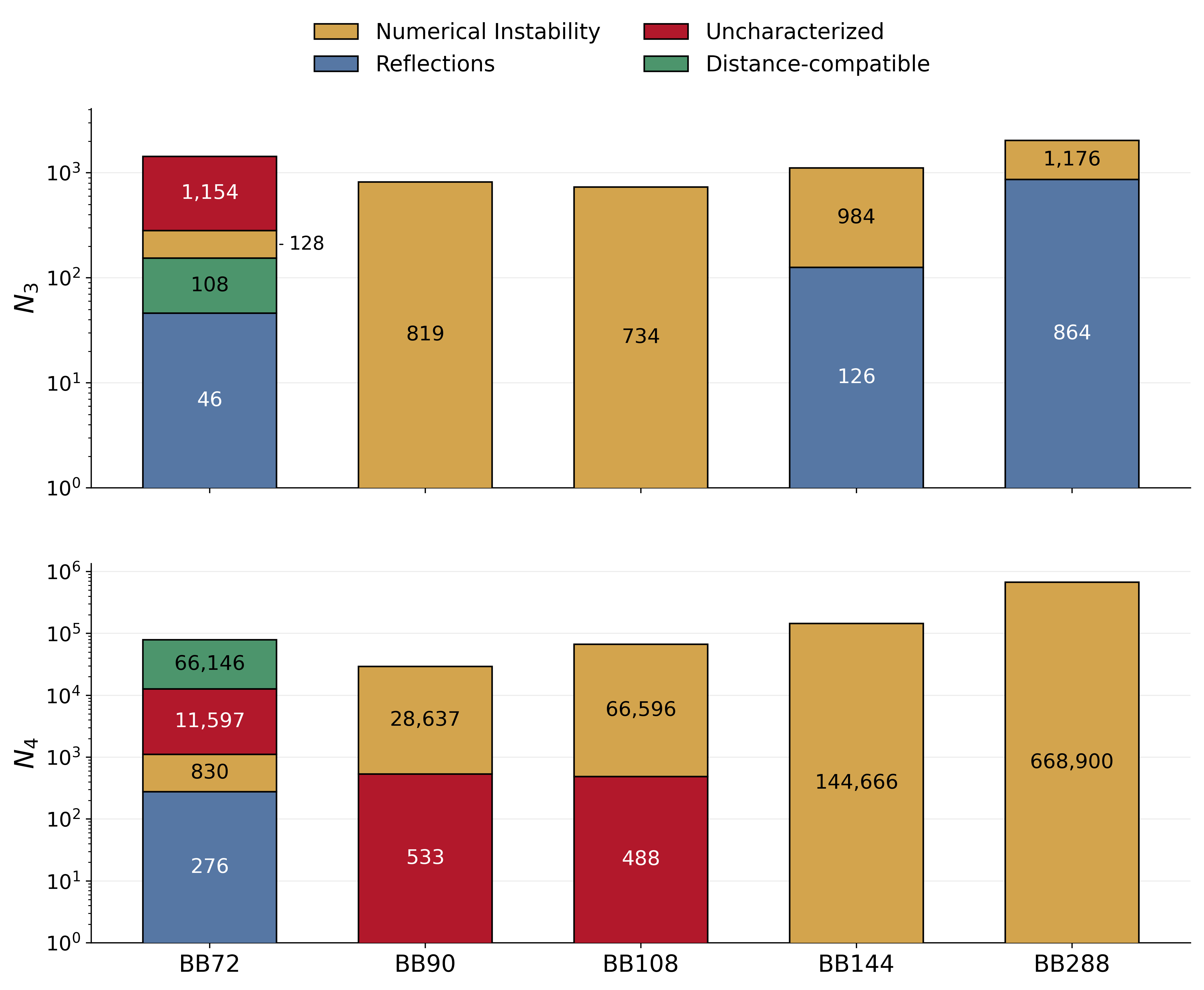}
    \caption{
          Cause of decoding failure for errors of weight $w=3,4$. The vertical bars (log scale) represent the counts obtained from exhaustive simulations. Colors indicate the identified causes of failure, stacked from smallest to largest.
    }\label{figure:counters_weight_34}
\end{figure}

The correctability of weight-$4$ errors suggest that the weight-$3$ failures follow a structured pattern. Investigating the low-weight problematic errors more deeply, we discovered that they were indeed caused by a specific dynamic, relevant for the codes \BB{72}, \BB{144} and \BB{288}. This mechanism is caused by special symmetries of the decoding graph, that we named reflections. Reflections are described in better detail in \cref{rare_events_analysis:reflections}. The contribution of reflections to the counts $N_3$ and $N_4$ is depicted in \cref{figure:counters_weight_34}. The picture additionally shows the component due to numerical instability, removed from the results in \cref{table:min_weight_simulations}. For the distance-$6$ code \BB{72}, the graph additionally displays the contribution of decoding failures which are compatible with the error correcting capabilities of the code. More precisely, the distance-compatible contribution contains an error $x$ if and only if $x\in \mathcal{F}$ and the solution $\tilde{x}$ proposed by the decoder is such that $\Hw{\tilde{x}} \leq \Hw{x}$. \\
Remarkably, our analysis completely characterizes the weight-$3$ failures for all codes, excluding \BB{72}.
\section{Reflections}
Our numerical results show that BP OSD decoder fails for some weight-$3$ errors for codes \BB{72}, \BB{144} and \BB{288}. While this is expected for the distance-$6$ code \BB{72}, this behavior does not align with the potential of \BB{144} and \BB{288}. We therefore analyzed the weight $3$-errors of the latter two code more precisely, and discovered that the support of a weight-$3$ failure is always contained in the support of an $X$-check of the code. The limitations of BP with these configurations are described in the literature by the concepts of split-beliefs~\cite{crigerMultipathSummationDecoding2018} and, more specifically, symmetric stabilizers~\cite{raveendranTrappingSetsQuantum2021}. Despite our errors are related to this mechanism, symmetric stabilizers alone are not sufficient to fully characterize our error patterns. The identified failures admit instead a more accurate description in terms of automorphisms of the Tanner graph, and they satisfy a precise list of properties that we grouped under the mathematical definition of \emph{reflection}. We numerically discovered that this precise pattern was not only able to explain the remaining weight-$3$ failures for \BB{144} and \BB{288}, but also a portion of the decoding failures for the code \BB{72}.

The remainder of this section is organized as follows. \Cref{rare_events_analysis:reflections} contains the definition and mathematical discussion of the reflection pattern. In \cref{subsec:symmetric_stabilizers}, we compare reflections with the concept of symmetric stabilizer~\cite{raveendranTrappingSetsQuantum2021}, highlighting the differences of the two definitions. \Cref{subsec:OSD_reflections} qualitatively explains why reflection errors lead to decoding failure even with OSD post-processing. Lastly, in \cref{subsec:decoding_variations} we demonstrate that reflections can be tackled by introducing targeted modifications of the decoder. We propose two variants of the decoder which achieve a substantial improvement in terms of logical error rate for the code \BB{144}. Both the proposed variations maintain the efficiency of (parallel) BP, and do not require OSD post-processing.

\subsection{Reflections and proof of BP Nonconvergence}\label{rare_events_analysis:reflections}
Loosely speaking, reflections are special symmetries of the Tanner graph that fix particular error syndromes. We formalize the intuitive concept of symmetry in \cref{def:symmetries}. This definition, more general than the notion of reflection, is motivated by \cref{lemma:symmetric_BP_evolution}, which describes how symmetries induce regularities in BP's hard decision. Next, in \cref{def:syndrome_admitting_reflection}, we give the additional properties that specify a reflection, and later prove that BP can not converge in presence of reflections (\cref{thm:BP_does_never_converge}). Finally, we report the structure of the found reflections for codes \BB{72}, \BB{144} and \BB{288}, which admit a representation in closed form.

In the following, we use the notion of code automorphism. Intuitively, a code automorphism is a permutation of variables and checks that preserves the structure of the code. For the purposes of this work, it is convenient to formulate this notion in graph-theoretic terms using the Tanner graph representation of a code.
\begin{definition}[Automorphism]
Let $\mathcal{T}=(V,C,E)$ be the Tanner graph of a classical binary code. Consider a couple of permutations $\sigma=(\sigma_V,\sigma_C)$, where $\sigma_V:V\to V$ and $\sigma_C:C\to C$. We say that $\sigma$ is an automorphism of $\mathcal{T}$ if the couple $(\sigma_V,\sigma_C)$ preserves the edge set $E$, i.e.,
\begin{equation*}
(\sigma_V(v),\sigma_C(c))\in E
\quad\text{if and only if}\quad
(v,c)\in E.
\end{equation*}
\end{definition}
We note that this definition differs from the notion of code automorphism commonly adopted in coding theory and quantum error correction. Nevertheless, it is straightforward to show that our definition is equivalent to the definition of a classical code automorphism given in Ref.~\cite{breuckmannFoldTransversalCliffordGates2024}. For the purposes of our discussion, we favor the graph-theoretic formulation, as it more naturally connects to the Tanner-graph structure on which BP operates.

For convenience, we use the loose notation $\sigma(v)=\sigma_V(v)$ and $\sigma(c)=\sigma_C(c)$ for $v\in V$ and $c\in C$, respectively. We also write
\begin{equation*}
    \sigma(x) = (x_{\sigma(v)})_{v \in V} \in \mathbb{F}_2^V\; \text{ and }\; \sigma(s) = (s_{\sigma(c)})_{c \in C} \in \mathbb{F}_2^C
\end{equation*}
and
\begin{equation*}
    \sigma(s) = (s_{\sigma(c)})_{c \in C} \in \mathbb{F}_2^C
\end{equation*}
to denote the permutations induced by $\sigma$ on the coordinates of errors and syndromes, respectively.

We begin with the formalization of the intuitive concept of symmetry.
\begin{definition}[Symmetry]\label{def:symmetries}
    We say that a Tanner graph automorphism $\sigma$ is a \emph{symmetry} for the syndrome $s$ if $\sigma(s)=s$.
\end{definition}
The symmetry $\sigma$ induces in turn a symmetric behavior in BP decoding. This is formalized in the following lemma. 
\begin{lemma}\label{lemma:symmetric_BP_evolution}
    Suppose $\sigma$ is a symmetry for syndrome $s$. Then, at every iteration of (parallel) BP, the hard decision $\tilde{x}$ satisfies $\sigma(\tilde{x}) = \tilde{x}$.
\end{lemma}
Despite the lemma focuses on the regularities of BP's hard decision, the ultimate cause of these regularities is the symmetric behavior of the underlying network of beliefs. This is made evident in the proof of the lemma (see Appendix \ref{lemma_proof:symmetric_BP_evolution}).

We remark that the results of \cref{lemma:symmetric_BP_evolution} and the following \cref{thm:BP_does_never_converge} hold for standard parallel BP, but they do not apply to serial or alternative BP scheduling~\cite{roffeDecodingQuantumLowdensity2020}. Even though our work always considers parallel BP decoding, we chose to clarify this in the two statements.

The dynamic described in \cref{lemma:symmetric_BP_evolution} is not harmful for BP decoding in general. However, the induced symmetry becomes problematic if the automorphism $\sigma$ satisfies the additional assumptions listed in the next definition.
\begin{definition}[Reflection]\label{def:syndrome_admitting_reflection}
    Suppose $\sigma$ is a symmetry for the syndrome $s$. We say that $\sigma$ is a reflection with fixed point $c\in\supp(s)$ if:
    \begin{enumerate}
        \item $\sigma$ is self-inverse, i.e., $\sigma \circ \sigma=\mathrm{id}$, \label{self-inverse}
        \item $\sigma(v) \neq v$ for every variable node $v\in V$, \label{no_fixed-qubits}
        \item $\sigma(c) = c$.\label{c_in_axis}
    \end{enumerate}
\end{definition}
The introduction of reflections with fixed point, that we call reflections for simplicity, is motivated by \cref{thm:BP_does_never_converge}. The theorem shows that the listed assumptions are sufficient to prove the non-convergence of BP when decoding this special pattern. The reason behind the name reflection comes instead from a geometric intuition. In basic geometry, reflections are self-inverse transformations which fix a \emph{reflection axis}. In our definition, the automorphism $\sigma$ is indeed a self-inverse function (property \ref{self-inverse}), with a nonempty set of invariant points $A =\{u \in V\cup C \mid \sigma(u) = u\}$. We think of $A$ as the reflection axis of $\sigma$. In particular, the definition specifies that $A \subseteq C$ (property \ref{no_fixed-qubits}) and $c \in A$ (property \ref{c_in_axis}). These assumptions are central in the proof of \cref{thm:BP_does_never_converge}, where we demonstrate that the linear recurrence specified by the check $c$ will never be satisfied by BP's hard decision. 
\begin{theorem}\label{thm:BP_does_never_converge}
    Let $s$ be a syndrome that admits a reflection $\sigma$ with fixed point $c \in \supp(s)$. \\
    Then, (parallel) BP decoding of $s$ never converges.
\end{theorem}
\begin{proof}
    Let $c$ be the fixed point specified in the statement. For every $v \in N(c)$, consider the set
    \[
        O_v = \{v, \;\sigma(v), \;\sigma \circ\sigma(v),  \;\sigma \circ\sigma \circ\sigma(v)\dots\}
    \]
    obtained by repeatedly applying the automorphism $\sigma$ to $v$. Since $c$ is a fixed point for the reflection $\sigma$, and $\sigma$ is a graph automorphism, it follows that $O_v \subseteq N(c)$. Moreover, given the assumption \ref{no_fixed-qubits}, $v$ and $\sigma(v)$ are distinct, and therefore $|O_v|\geq 2$. Additionally, \ref{self-inverse} guarantees that $\sigma \circ \sigma(v) = v$, which implies $|O_v|\leq2$ and therefore $O_v$ contains precisely two elements  $O_v = \{v, \sigma(v)\}$.
    We conclude that $N(c)$ can be written as the union of pairwise disjoint sets 
    \begin{equation*}
        N(c) = \bigcup _ {v\in S} \{v, \sigma(v)\}
    \end{equation*}
    for some $S \subset N(c)$ of cardinality $\frac{|N(c)|}{2}$.
    Consider now the hard decision $\tilde{x}$ produced by the parallel BP decoder at any iteration of the algorithm. The candidate $\tilde{x}$ does not satisfy the parity-check constraint $\bigoplus_{v \in N(c)} \tilde{x}_v = s_c = 1$, because
    \begin{equation*}
        \bigoplus_{v \in N(c)} \tilde{x}_v = \bigoplus_{v \in S} (\tilde{x}_v \oplus \tilde{x}_{\sigma(v)}) = \bigoplus_{v \in S} 2 \, \tilde{x}_v = 0 \mod 2,
    \end{equation*}
    where the second equality holds thanks to \cref{lemma:symmetric_BP_evolution}. As a consequence, the hard decision is never compatible with the syndrome $s$, and the decoder never converges.
\end{proof}

In our analysis, we interpret \cref{lemma:symmetric_BP_evolution} and \cref{thm:BP_does_never_converge} as the mechanism underlying the BP decoding failure in the presence of reflections. Specifically, in \cref{figure:counters_weight_34}, we count as reflection the errors whose syndrome $s$ satisfies the properties listed in \cref{def:syndrome_admitting_reflection} for some symmetry $\sigma$. In order to identify $\sigma$, we evaluated which variables of the estimated soft decision share the same value across multiple iterations, and paired them to define a candidate function $\sigma$. We successively verified that $\sigma$ is a Tanner graph automorphism and a reflection with fixed point for the given syndrome. Considering a particular BB code, we discovered that all the reflection errors share the same automorphism $\sigma$, which is unique up to translation along the lattice. Moreover, $\sigma$ can be represented in closed form, that we report in \cref{table:reflections}. 
\begin{table}[htbp]
    \centering
    \begin{tabular}{ccc}
        \toprule
        Code & \hspace{1 cm}  & $\sigma|_L [(L, x^a \, y^b)]$ \\
        \midrule
        \BB{72} &  & $(R,x^b \, y^a)$ \\
        \BB{90} &   & - \\
        \BB{108} &   & - \\
        \BB{144} &   & $(R,x^{9a+2b+10} \, y^{2a+3b})$ \\
        \BB{288} &   & $(R,x^{9a+2b+10} \, y^{2a+3b+6})$ \\
        \bottomrule
    \end{tabular}
    \caption{Tanner graph automorphisms $\sigma$ underlying the reflection failures. For each code, $\sigma$ maps $L$-qubits to $R$-qubits and vice versa. The table reports a closed form for the restriction of $\sigma$ to $L$-qubits $\sigma|_L : L \to R$. The action on $R$-qubits is given by the inverse map. For the notation used to denote qubits and checks in BB codes, see appendix \ref{sec:definitions}.}
    \label{table:reflections}
\end{table}
The structure of $\sigma$ is particularly simple to visualize for the code \BB{72}, where it corresponds to a diagonal reflection of the lattice (see \cref{fig:BB72_iterations}).

\subsection{Relation to previous work: Symmetric Stabilizers}\label{subsec:symmetric_stabilizers}
In this section, we specify analogies and differences between the reflection pattern and previous work. In Ref.~\cite{raveendranTrappingSetsQuantum2021}, the authors observed that local symmetries in the syndrome can be harmful for message-passing decoders, and they introduced the definition of symmetric stabilizer. The proof of \cref{lemma:symmetric_BP_evolution} is based on their argument motivating the failure mechanism of symmetric stabilizers. Additionally, considering a syndrome $s = x H^T$ which admits a reflection $\sigma$, $x \,\oplus\,\sigma(x)$ is either a stabilizer or a logical operator of the code. In the first case, it satisfies the definition of symmetric stabilizer \footnote{This is evident considering the partition $\supp(x \,\oplus\,\sigma(x)) =\supp(x) \cup \supp(\sigma(x)) $ in the role of the partition of the symmetric stabilizer definition~\cite{raveendranTrappingSetsQuantum2021}}. 
However, a reflection is not locally constrained within the symmetric stabilizer support, but extends globally along the Tanner graph. On the contrary, the symmetric stabilizer characterizes the dynamics locally within the stabilizer support, and relies on the \emph{isolation assumption}~\cite{planjeryIterativeDecodingBelief2010a, chytasEnhancedMinSumDecoding2025} to mediate the interaction with the rest of the graph. \\
\Cref{def:syndrome_admitting_reflection} is therefore a stronger requirement compared to symmetric stabilizers. \Cref{thm:BP_does_never_converge}, utilizes these additional requests to prove the decoder non-convergence without general assumptions on the decoding dynamics. 

We observe that the difference between the two definitions is not merely an alternative description of the same dynamics, but it is instead necessary to explain the outcome of our simulations.  We detail two arguments that support this necessity in the following paragraphs.

First, since the Bivariate Bicycle codes listed in~\cite{bravyiHighthresholdLowoverheadFaulttolerant2024} share the same local check structure, a discussion in terms of symmetric stabilizers would be identical for every code. Conversely, the behavior we observed is dependent on the code. Indeed, only the codes \BB{72}, \BB{144} and \BB{288} exhibit weight-$3$ failures. For each of these cases, the local symmetry extends to the whole decoding graph. In contrast, we do not observe the same failures for codes \BB{90} and \BB{108} for the numerically stable version of the algorithm. Additional data from our simulation demonstrates that BP converges for every weight-$3$ error for these two codes, and therefore this behavior is not caused by the OSD post-processing. Given this numerical evidence, and considered  \cref{thm:BP_does_never_converge}, we conjecture that BP properly decodes all the weight-$3$ errors for \BB{90} and \BB{108} because their Tanner graphs do not admit (global) reflections. 

\begin{figure}
    \centering
    \includegraphics[width=.7\linewidth]{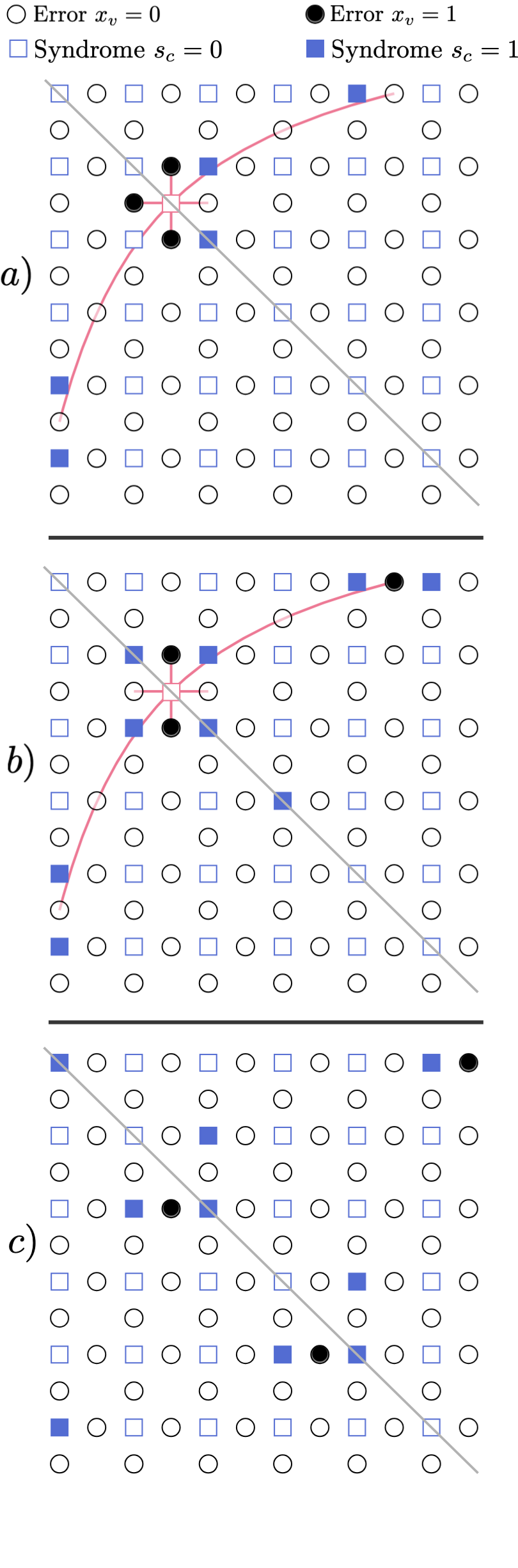}
    \caption{
          Three examples of errors that highlight the difference between reflections and symmetric stabilizers, code \BB{72}, where $\sigma$ corresponds to a reflection along the diagonal axis (gray line). \textbf{a)} An error (black dots) contained in the support of an $X$-check (red), whose syndrome is not invariant under the action of the reflection $\sigma$. We did not abserve failures of this type in our experiments, despite they are predicted by the symmetric stabilizer formalism. \textbf{b)} An error contained in the support of an $X$-check, with syndrome invariant under $\sigma$. \textbf{c)} An error contained in the support of a weight-$6$ logical-$X$ operator of the code. The syndrome is invariant under $\sigma$. All the logical failures counted in the reflection contribution in \cref{figure:counters_weight_34} come from errors of the type b) and c). 
    }\label{fig:error_examples}
\end{figure}

Another argument depends instead on the particular triggering patterns of symmetric stabilizer and reflections. Due to the structural symmetries of $\mathcal{T}'$, the symmetric-stabilizer formalism predicts that BP fails to converge for every weight-$3$ error $x$ with $\supp(x)\subseteq\supp(c_X)$. Conversely, the reflection failures observed in our simulations have different properties, that can be described as two chategories of errors. The first forms a strictly smaller subset of the errors predicted by the symmetric stabilizer formalism, containing errors whose syndromes are invariant under the action of the automorphisms $\sigma$ reported in \cref{table:reflections}. The second class contains instead errors $x$ such that $x \,\oplus\, \sigma(x)$ constitutes a logical operator of the code. Examples supporting this discussion are depicted in \cref{fig:error_examples}. 

Finally, we remark that the dynamic induced by symmetric stabilizers and reflections would be indistinguishable if the maximum number of BP iterations $T$ is small ($T < 100$). Setting a small $T$ would indeed prevent the global graph asymmetries to contribute in the soft decision estimation, and induce a stronger dependence from the local pattern.

\newpage
\subsection{OSD Post-Processing of Reflections}\label{subsec:OSD_reflections}
\begin{figure}
    \centering
    \includegraphics[width=1\linewidth]{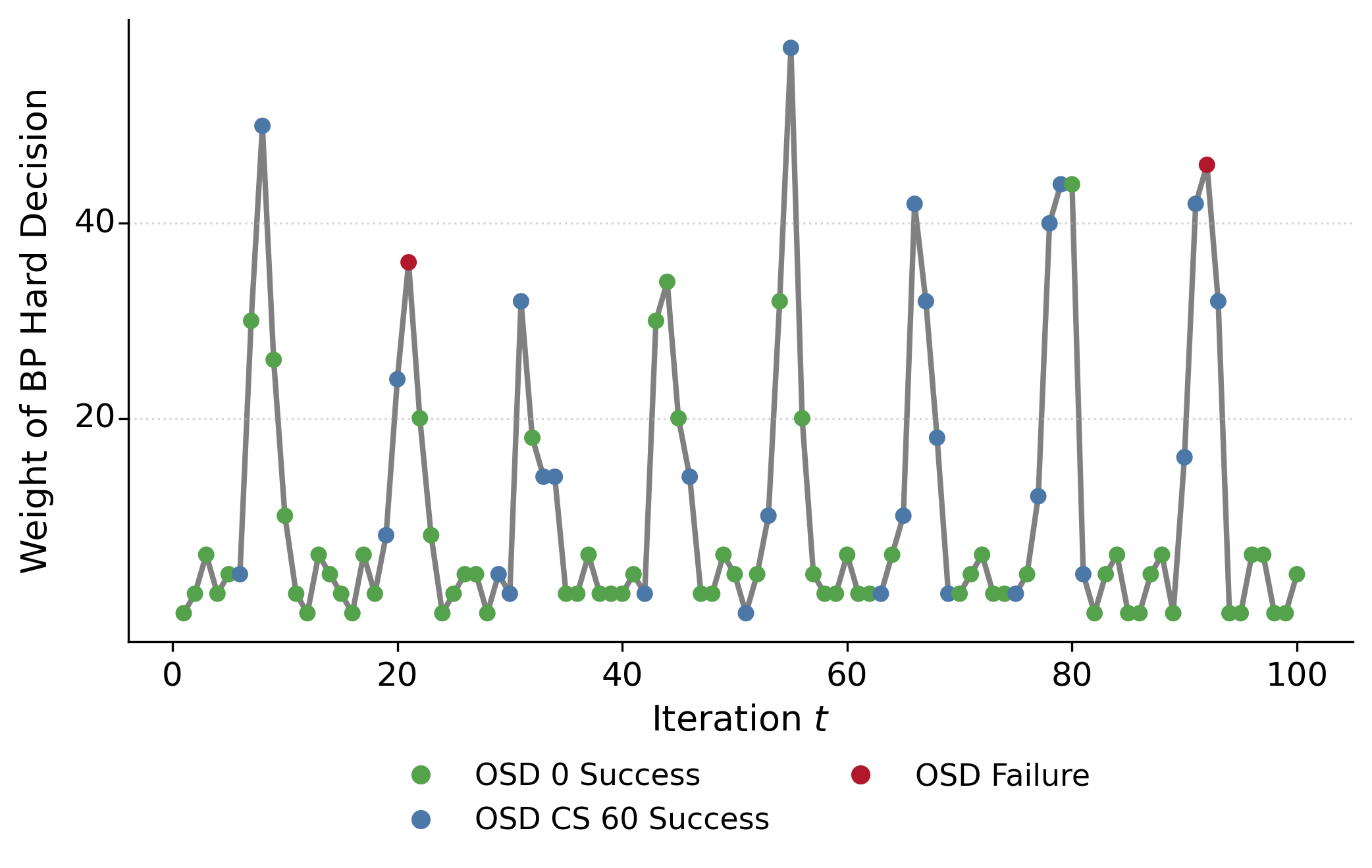}
    \caption{
    Weight of BP's hard decision and effectiveness of OSD in decoding a reflection error, code \BB{144} code. The graph shows the weight of the hard decision as a function of the iteration counter $t$. The marker colors denote success or failure of the OSD post-processing when BP is halted at the corresponding iteration ($T = t$). Green markers represent successful decoding using OSD 0. Blue markers denote failure of OSD 0 but success of OSD CS 60. Red markers indicate failure of both OSD 0 and OSD CS 60.
    }
    \label{fig:decoder_evolution}
\end{figure}

The inaccuracy of BP for quantum codes is a known issue, and the role of the OSD post-processing is precisely to tackle BP's limitations. It is therefore natural to investigate why the OSD post-processing is not effective in overcoming the identified reflection mechanism.
In this section, we qualitatively argue that the insuccess of OSD decoding depends in this case on the quality of BP's hard decision. Specifically, \cref{fig:decoder_evolution} shows the evolution of the BP hard decision weight during the decoding of a reflection error. The curve exhibits recurring sequences of growth phases and contraction regimes, although without strict periodicity. At the peaks, the hard decision reaches weights significantly larger ($>30$) than that of the original error ($3$).

The graph shows that the effectiveness of OSD post-processing depends strongly on the iteration at which BP is terminated. If the maximum iteration count $T$ is reached during a high-weight regime, OSD 0 is often unable to recover the correct solution (blue and red markers in \cref{fig:decoder_evolution}). We tested the more robust OSD CS 60 variant of the post-processing procedure~\cite{roffeDecodingQuantumLowdensity2020}. Although it provides a substantial improvement over OSD 0 (blue markers), OSD CS 60 still fails to recover a correct solution for some BP stopping iterations $t$ (red markers).

We conclude that OSD improves in BP decoding of reflections, but our analysis also indicates that OSD does not guarantee their complete correctability. More precisely, our qualitative discussion provides evidence that the OSD limitations in decoding reflections stems from unreliable hard decision. As a consequence, we suggest that solving the reflection pattern will require a modification of the initial BP procedure, or a post-processing step which tolerates misleading soft or hard decisions produced by BP. In our experiments with custom decoders (see \cref{subsec:decoding_variations}), we explored the first strategy. To explore the second approach, a possible direction for future work is to employ an ensemble of BP OSD decoders configured with different values of $T$. This can be implemented in practice by applying OSD to multiple versions of the hard decision obtained at different BP iterations.

\newpage
\subsection{Improved Decoding}\label{subsec:decoding_variations}
Here we describe additional numerical experiments which demonstrate that the impact of the found low weight failures can be mitigated. We devised two custom versions of Belief Propagation that we called \emph{Noisy BP} and \emph{Detect-Hold BP}. First, we briefly describe the two decoders, whose more detailed description can be found separately in boxes \ref{box:noisy_bp} and \ref{box:detect_hold_bp}. In the boxes, the blue text highlights the difference from the standard BP algorithm. Next, we report the performance of the decoders in terms of estimated logical error rate.

Both Noisy BP and Detect-Hold BP are modifications of Belief Propagation, and do not rely on OSD post-processing. The implementations of the two decoders have been developed as modifications of the software in ref.~\cite{roffeDecodingQuantumLowdensity2020}. The two decoders also include the numerical stability improvement described in \cref{numerical_instability}.

Noisy BP is inspired by the well known perturbation strategy~\cite{poulinIterativeDecodingSparse2008}. Intuitively, this approach attempts to break the symmetries underlying the reflection errors by randomly perturbing BP's beliefs. In our particular decoder, the perturbation term is added in \cref{noisy_BP_update_rule} and bounded to $[-\epsilon, + \epsilon]$, where $\epsilon \in \mathbb{R}$ is a chosen parameter.\\

\begin{algorithmbox}\label{box:noisy_bp}
\textbf{Noisy BP}\\\\
With the same notation of the BP algorithm \ref{box:BP}, let $t\in\mathbb{N}$ be the iteration counter, let $m^t_{v\to c}$ and $m^t_{c\to v}$ be the variable-to-check and check-to-variable messages respectively. Denote $\LLR(\pi) = \ln(\frac{1-\pi}{\pi})$ and $f(\pi) = \tanh(\pi/2)$ for $\pi\in\mathbb{R}$. \\

At $t=0$, $m^0_{v\to c} = \LLR(p)$, where $p$ is the considered physical error rate. Check-to-bit messages $m^t_{c\to v}$ are not defined for $t=0$. \\
For $t \geq 1$, the algorithm performs the following steps.
\begin{enumerate}
    \item For every edge $(c,v) \in E$, compute the messages:
        \begin{align}
            m^t_{c\to v} &= (-1)^{s_c} f^{-1}\left[\prod_{v'\in N(c)\setminus \{v\}} f(m^{t-1}_{v' \to c})\right] ,\\
            m^t_{v \to c} &= \highlight{ \epsilon^t_{c\to v} +} \LLR(p) + \sum_{c'\in N(v)\setminus \{c\}} m^t_{c' \to v},\label{noisy_BP_update_rule}
        \end{align}
    \highlight{where the perturbation term $\epsilon^t_{c\to v}$ is uniformly drawn in the interval $[-\epsilon, +\epsilon]$.}
    \item Execute steps \ref{BP_step_2} and \ref{BP_step_3} of the standard BP algorithm.
    \highlight{\item Replace potential numerical errors in messages $m^t_{v\to c}$ with $\LLR(\frac{1}{2}) = 0$ to improve num. stability.}
\end{enumerate}
If a maximum iteration counter $T$ is reached, the decoder terminates outputting a failure status.\\

\textbf{Selected Parameters}\\
In our numerical experiment in \cref{fig:decoder_performance}, we selected \\$p=10^{-4}$, \highlight{$\epsilon =10^{-5}$}, $T=500$. 
\end{algorithmbox}

Detect-Hold BP is instead an attempt to enforce an asymmetry more actively. The decoder is a variation of BP specialized in identifying and tackling the mechanism induced by errors of type b) shown in \cref{fig:error_examples}. The identification phase (Detect) is based on the oscillatory pattern described in Theorem 1 of ref.~\cite{chytasEnhancedMinSumDecoding2025} in the context of symmetric stabilizers. After detection, the algorithm freezes the value of selected variables within the stabilizer support (Hold). Detect-Hold BP requires two additional parameters: a \emph{waiting time} $T_w \in \mathbb{N}$ and a \emph{hold-parameter} $\pi_h \in \mathbb{R}$.\\

\begin{algorithmbox}\label{box:detect_hold_bp}
\textbf{Detect-Hold BP}\\\\
Denote with $\mathcal{T}_X = (Q, C_X, E_X)$ and $\mathcal{T}_Z = (Q, C_Z, E_Z)$ the Tanner graphs of a quantum CSS code. Detect-Hold BP is a variation of BP applied to the Tanner graph $\mathcal{T}_Z$. Using the same notation of the BP algorithm \ref{box:BP} with $\mathcal{T} = \mathcal{T}_Z$, let $t\in\mathbb{N}$ be the iteration counter, let $m^t_{v\to c}$ and $m^t_{c\to v}$ be the variable-to-check and check-to-variable messages respectively. Denote $\LLR(\pi) = \ln(\frac{1-\pi}{\pi})$ and $f(\pi) = \tanh(\pi / 2)$ for $\pi\in\mathbb{R}$. \\

At $t=0$, $m^0_{v\to c} = \LLR(p)$, where $p$ is the considered physical error rate. Check-to-bit messages $m^t_{c\to v}$ are not defined for $t=0$. \\
For $t \geq 1$, the algorithm performs the following steps.
\begin{enumerate}
    \item Execute steps \ref{BP_step_1}, \ref{BP_step_2} and \ref{BP_step_3} of the standard BP algorithm. Denote respectively with $d^t$ and $\tilde{x}^t$ the obtained soft and hard decision vectors at the current iteration $t$.
    
    \highlight{\item Replace potential numerical errors in messages $m^t_{v\to c}$ with $\LLR(\frac{1}{2}) = 0$ to improve num. stability.}
    
    \highlight{\item For every $X$-check $c_x \in C_X$, consider $v$ belonging its neighborhood $N(c_x)\subseteq Q$ and test $\tilde{x}^t_v = \tilde{x}^{t-1}_v \oplus 1$. If equality is satisfied for every $v\in N(c_x)$, select $v' = \argmax_{v\in N(c_x)} |d_v^t -d_v^{t-1}| $ and hold fix the messages $$m^{t'}_{v'\to c} = \pi_h$$ for every $t' \in [t, t + T_w]$ and $c\in N(v')\subseteq C_Z$.}
    
    \end{enumerate}
If a maximum iteration counter $T$ is reached, the decoder terminates outputting a failure status.\\

\textbf{Selected Parameters}\\
In our numerical experiment in \cref{fig:decoder_performance}, we selected \\$p=10^{-4}$, $T=500$, \highlight{$T_w = 30$, $\pi_h = 100$}. 
\end{algorithmbox}

\begin{figure}
    \centering
    \includegraphics[width=1\linewidth]{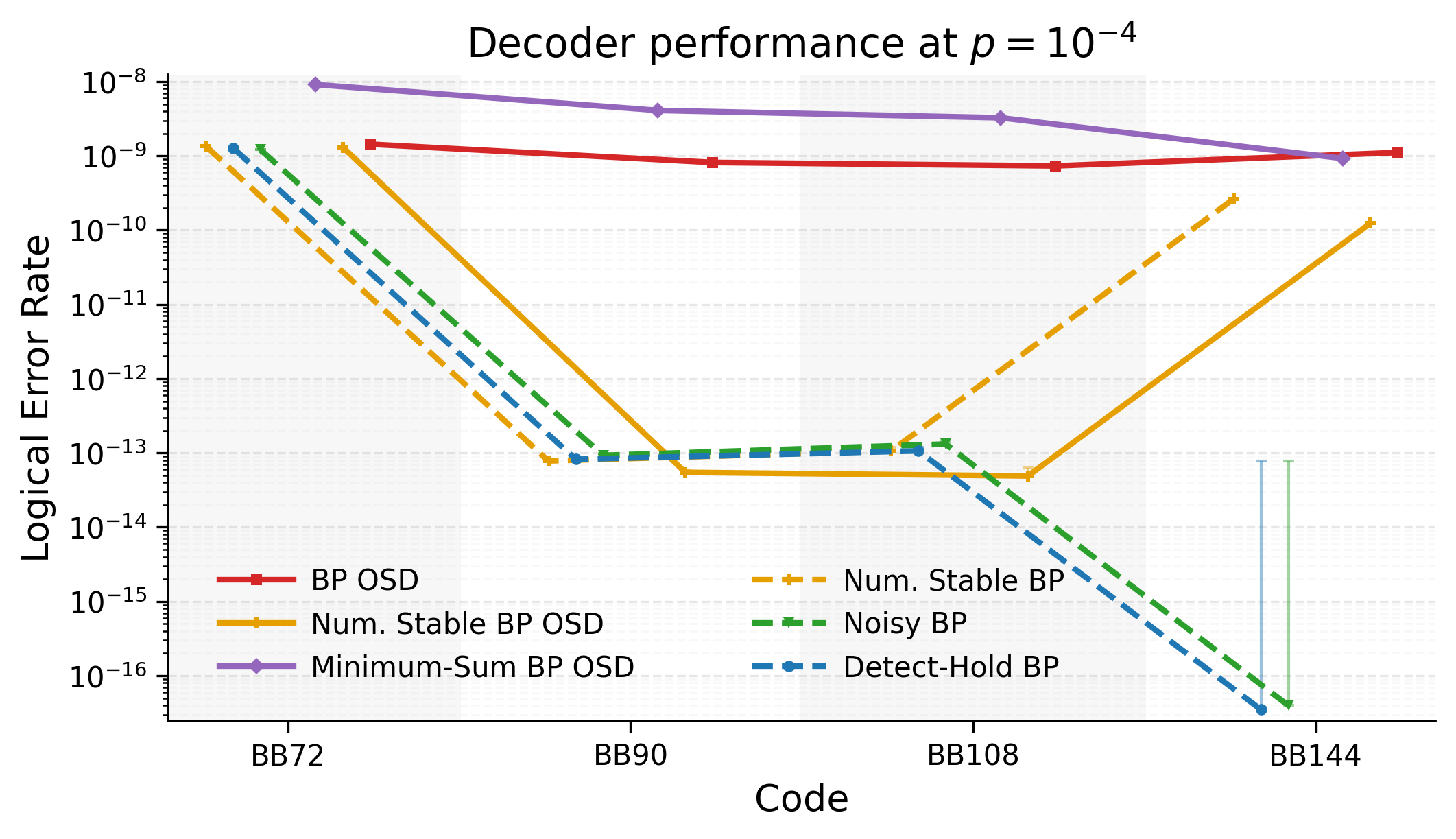}
    \caption{
    Decoder performance comparison across different BB codes at physical error rate $p = 10^{-4}$. Solid (dashed) lines correspond to BP-based decoders including (excluding) OSD post-processing. For reference, the graph includes the Minimum-Sum variant of BP OSD~\cite{roffeDecodingQuantumLowdensity2020} (purple). The upper error bars indicate bounds on the contribution of higher-weight errors, as described in \cref{eq:LER_estimate}. Results for \BB{288} are not reported because the bound is insufficiently tight to allow meaningful conclusions.
    }
    \label{fig:decoder_performance}
\end{figure}

To assess the performance of the introduced decoders, we repeated the exhaustive numerical simulations described in \cref{subsec:methods} for Noisy BP and Detect-Hold BP, also including other decoder variants. The simulation results, expressed as logical error rates, are shown in \cref{fig:decoder_performance}.

The comparison shows that Noisy BP and Detect-Hold BP achieve performance close to that of numerically stable BP for codes \BB{72}, \BB{90}, and \BB{108}. This result, consistent with reflections being a negligible source of decoding failures for these codes, suggests that our modifications of BP do not substantially alter its performance in decoding general errors. Numerically stable BP OSD performs better, illustrating the versatility of the post-processor, which can correct a broader set of failures beyond those associated with reflections.

For code \BB{144}, by contrast, Noisy BP and Detect-Hold BP exhibit a substantial improvement, supporting the interpretation that reflections are a major source of logical failures in this case. Remarkably, the two modified versions outperform numerically stable BP OSD even without post-processing, improving decoding performance while avoiding the additional computational cost of OSD.

Lastly, we remark that more sophisticated and more thoroughly analyzed BP decoders based on principles similar to those of Noisy BP and Detect-Hold BP exist~\cite{mullerImprovedBeliefPropagation2025}. The Detect-Hold strategy, in particular, closely resembles the BP OTS decoder of~\cite{chytasEnhancedMessagePassingDecoding2024} and guided decimation~\cite{yaoBeliefPropagationDecoding2024}. Additionally, other decoding techniques that we did not test in our investigations might effectively mitigate the observed dynamics, such as stabilizer inactivation~\cite{crestStabilizerInactivationMessagePassing2022} or the use of overcomplete parity-check matrices~\cite{miaoNeuralBeliefPropagation2023}. Serial BP scheduling~\cite{roffeDecodingQuantumLowdensity2020} also represents a potential solution to failures induced by symmetries.

Overall, our experiments do not aim to propose new general-purpose decoders. Instead, they demonstrate that the reflection mechanism can have a major impact on decoding performance, while also admitting practical remedies that preserve the parallel structure of belief propagation.

\section{Concluding remarks}

In this work, we numerically investigated low-weight failures of the BP OSD decoder under bit-flip noise at low physical error rates. We characterized its suboptimal performance and identified mechanisms responsible for these failures.

Whereas previous work explains such events in terms of local patterns, we found that the failures explained here depend on a stronger symmetry extending across the decoding graph. This suggests that BP is robust against many of the local patterns previously considered problematic, and that it may benefit from the nonuniform fault probabilities or defects present in realistic systems. We call these symmetries reflections, formally characterized their effect on BP dynamics, and numerically identified the actual symmetries in the studied codes.

Our analysis provides evidence that, in the presence of reflections, the performance of off-the-shelf BP OSD is governed more by check weight than by code distance. This highlights the need to address these error patterns. Through experiments with custom decoder variants, we showed that these failure mechanisms can be mitigated efficiently. In particular, BP variants that break reflection symmetries during decoding are more effective than post-processing techniques, because the soft information produced in the presence of reflections is unreliable.

As an additional finding, we identified a code-decoder combination whose failure spectrum cannot be reliably studied using the Splitting method~\cite{bravyiSimulationRareEvents2013, mayerRareEventSimulation2025, beverlandFailFastTechniques2025}. Our work also led to the detection and resolution~\cite{ldpc_NaN_github_issue} of a numerical issue in a widely used implementation of BP OSD~\cite{roffeDecodingQuantumLowdensity2020}. Given the implementation's wide adoption, this finding may be relevant to other low error rate simulations in the QEC community.

Overall, these results clarify the mechanisms limiting BP OSD at low physical error rates under code-capacity noise. An important direction for future work is to determine whether these mechanisms remain relevant under circuit-level noise, where additional correlations and measurement errors may change the dominant sources of decoding failure.


\section*{Acknowledgments}
We thank T. Yoder and M. Carroll for helpful discussions on the splitting method, K. Hansenne on low-weight error analysis, J. Roffe on BP OSD, and D. Chitas and M. Pacenti on symmetric stabilizers and BP failures.
BJB is grateful to the Niels Bohr Institute at the University of Copenhagen for their hospitality.
We acknowledge financial support from Danmarks Innovationsfond (IFD1003402609, FTQP) and VILLUM FONDEN Synergy (VIL78724). 
S.P. acknowledges funding from VILLUM FONDEN (MapQP, No. VIL60743), the European Research Council (ERC StG ASPEQT, No. 101221875), and Danmarks Innovationsfond
research grant No. 4356-00009B (HyperTenQ), and funding support from the NNF Quantum Computing Programme.

\section*{Author contributions}
All authors contributed to writing the manuscript and to the discussions that shaped the project. 
MCL performed the exhaustive numerical simulations. GF designed and wrote the code used in the numerical analysis, interpreted the numerical results, identified the reflection symmetries, and developed the custom decoder modifications. 
OpenAI Codex was used to write part of the software and assist with manuscript revision.

\section*{Data and code availability}
The data and code supporting the numerical results of this work are available on Zenodo~\cite{fregonaDealingLowweightErrors2026a}.

\bibliographystyle{apsrev4-2}
\bibliography{BB_BPOSD_paper.bib}

\appendix
\renewcommand{\thesubsection}{\thesection.\arabic{subsection}}
\renewcommand{\thesubsubsection}{\thesubsection.\arabic{subsubsection}}
 
\setcounter{secnumdepth}{2}
 
\setcounter{figure}{0}
\setcounter{table}{0}

\section{Definitions}\label{sec:definitions}

\subsection{Classical codes}

Given a vector space $V$ and a subspace $W\subseteq V$, let $W^\perp$ denote the orthogonal complement of $W$ with respect to the standard dot product. 

\begin{definition}[Classical code]
    A (binary) classical linear code $\mathcal{C}$ is a linear subspace of $\mathbb{F}_2^n$. A matrix $H \in\mathbb{F}_2^{m \times n}$ whose rows form a generating set of $\mathcal{C}^\perp$ is called a parity-check matrix of $\mathcal{C}$. 
\end{definition}
In coding theory, the vectors belonging to the code $\mathcal{C}$ represent the possible configurations of the encoded information. The parity-check matrix of a code can be used to test membership in $\mathcal{C}$ of a given vector $v\in\mathbb{F}_2^n$, indeed $v \in \mathcal{C}$ is equivalent to $v H^T = 0$.

For a given choice of parity-check matrix, the code admits an alternative representation in terms of the Tanner graph.
\begin{definition}[Tanner graph]
    Consider a classical linear code with parity-check matrix $H\in\mathbb{F}_2^{m \times n}$. Let $V$ and $C$ be the sets of column and row indices of $H$ respectively. The Tanner graph $\mathcal{T} = (V, C, E)$ associated to the matrix $H$ is a bipartite graph with vertex set $V \cup C$, where $(v,c)\in E$ if and only if $H_{c,v}=1$.
\end{definition}
In this context, the elements of $V$ are usually called \emph{variables}, while the elements of $C$ are called \emph{checks}.\\
For simplicity, we will also refer to $\mathcal{T}$ as the Tanner graph of $C$, despite $\mathcal{C}$ admits alternative Tanner graphs representations.

\subsection{Quantum CSS codes}
\begin{definition}[Quantum CSS code]
    A quantum CSS code is given by a pair of classical linear codes $\mathcal{C}_X, \mathcal{C}_Z \subseteq \mathbb{F}_2^n$ satisfying the \emph{commutation property} 
    \begin{equation*}
        \mathcal{C}_X^\perp \subseteq \mathcal{C}_Z.
    \end{equation*}
    The parameter $n$ is called \emph{number of physical qubits}, $k \df n - \dim(\mathcal{C}_X^\perp)- \dim(\mathcal{C}_Z^\perp)$ is called \emph{number of logical qubits}. The \emph{distance} of the code $d$ is the minimum weight of the vectors in $(\mathcal{C}_X \setminus \mathcal{C}_Z^\perp) \cup (\mathcal{C}_Z \setminus \mathcal{C}_X^\perp)$.
\end{definition}
The parameters of a quantum code are conventionally reported in square brackets $\llbracket n,k,d \rrbracket$.

The two classical codes $\mathcal{C}_X, \mathcal{C}_Z$ can be defined via two parity-check matrices $H_X, H_Z \in \mathbb{F}_2^{m \times n}$, where we here assume for simplicity that these two matrices have the same number of rows $m\in\mathbb{N}$. The commutation relation can be expressed in matrix form $H_X H_Z^T = 0$, meaning that the two matrices have orthogonal rows. \\
Interpreting $\mathcal{C_X}^\perp$ as the row span of $H_X$, the commutation relation becomes
\[
    \rowspan(H_X) = \mathcal{C}_X^\perp \subseteq \mathcal{C}_Z,
\]
and therefore the set of the rows of $H_X$ can be extended to a generating set of $\mathcal{C}_Z$. More precisely, the number of additional rows required by this procedure is 
\[
    \dim(\mathcal{C}_Z) - \dim(\mathcal{C}_X^\perp) = n - \dim(\mathcal{C}_Z^\perp) - \dim(\mathcal{C}_X^\perp) = k.
\]
Let $A_X \in \mathbb{F}_2^{k \times n}$ denote the matrix formed by the $k$ additional rows. The matrix $A_X$ is called \emph{logical-$X$ action matrix} of the quantum CSS code. Analogously, we can define a \emph{logical-$Z$ action matrix} $A_Z\in \mathbb{F}_2^{k \times n}$. Remarkably, $A_X$ and $A_Z$ are not unique.


Quantum CSS codes can be alternatively described in terms of Tanner graphs. Let $\mathcal{T}_X = (Q, C_X, E_X)$ be the Tanner graph associated to $H_X$ and $\mathcal{T}_Z = (Q, C_Z, E_Z)$ be the Tanner graph associated to $H_Z$. In this case, the set of variables of the two Tanner graphs $Q$ coincide, because the column indexes of both $H_X$ and $H_Z$ are the same, and it is called the set of qubits. The elements of $C_X$ and $C_Z$ are called $X$-checks and $Z$-checks respectively.
\begin{definition}[Tanner graph of a quantum CSS code]
    Let $\mathcal{C}_X, \mathcal{C}_Z \subseteq \mathbb{F}_2^n$ be a quantum CSS code, and suppose $H_X$ and $H_Z$ are parity-check matrices of $\mathcal{C}_X$ and $\mathcal{C}_Z$ respectively. Let $\mathcal{T}_X = (Q, C_X, E_X)$ and $\mathcal{T}_Z = (Q, C_Z, E_Z)$ be the associated Tanner graphs, with $Q = [1, \dots n]$. The vertices of $C_X$ are called \emph{$X$-checks}, the elements of $C_Z$ are called \emph{$Z$-checks}, and the vertices in $Q$ are called \emph{qubits}. 
\end{definition}

\subsection{Bivariate Bicycle codes}\label{BB_codes}

In our numerical simulations, we consider a class of CSS codes belonging to the Bivariate Bicycle (BB) family. In this section, we provide a concise description of BB codes sufficient for the purposes of this article. For a detailed treatment of their construction and properties, we refer the reader to ref.~\cite{bravyiHighthresholdLowoverheadFaulttolerant2024}.

BB codes are conveniently described through the structure of their Tanner graphs. Qubits and check nodes are labelled by bivariate polynomials. Let \(l,m\in\mathbb{N}\), and define the ring \(l,m\in\mathbb{N}\), and let $\mathcal{R}=\mathbb{F}_2[x,y]/\langle x^l+1, y^m+1\rangle$. 
Let 
\begin{equation}\label{BB_codes:M}
    \mathcal{M} = \{1,y,\ldots,y^{m-1},x,xy,\ldots,xy^{m-1},\ldots,x^{l-1}y^{m-1}\}
\end{equation}    
denote the set of monomials of $\mathcal{R}$.

The qubit set $Q$ consists of two disjoint copies of \(\mathcal{M}\). Following the notation of ref.~\cite{bravyiHighthresholdLowoverheadFaulttolerant2024}, we write $Q = \{L, R\} \times \mathcal{M}$, where $Q_L=\{L\}\times\mathcal{M}$,  $Q_R=\{R\}\times\mathcal{M}$, are referred to as the left and right qubits, respectively. Similarly, the \(X\)- and \(Z\)-check nodes are labelled by
$C_X=\{X\}\times\mathcal{M}$, $C_Z=\{Z\}\times\mathcal{M}$. Here $L$, $R$, $X$ and $Z$ are labels identifying the qubit and check types. 

The Tanner graph is determined by two polynomials
\[
A=A_1+A_2+A_3,
\qquad
B=B_1+B_2+B_3,
\]
where $A_1,A_2,A_3\in\mathcal{M}$ are distinct, and likewise $B_1,B_2,B_3\in\mathcal{M}$ are distinct.\\
The edge set $E_X$ consists of the edges
\[
\bigl((X,\alpha),(L,\alpha A_i)\bigr),
\qquad
\bigl((X,\alpha),(R,\alpha B_j)\bigr),
\]
for every $(X,\alpha)\in C_X$ and $i,j\in\{1,2,3\}$. \\
Similarly, the edge set $E_Z$ consists of
\[
\bigl((Z,\beta),(L,\beta B_j^{-1})\bigr),
\qquad
\bigl((Z,\beta),(R,\beta A_i^{-1})\bigr),
\]
for every $(Z,\beta)\in C_Z$ and $i,j\in\{1,2,3\}$.

The resulting Tanner graphs
\begin{equation}\label{BB_codes:Tanner_graph}
\mathcal{T}_X=(Q,C_X,E_X),
\qquad
\mathcal{T}_Z=(Q,C_Z,E_Z),
\end{equation}
define the corresponding BB code, as formalized below.
\begin{definition}
Let $l,m\in\mathbb{N}$, let $\mathcal{M}$ be as in \cref{BB_codes:M}, and let
\[
A=A_1+A_2+A_3,
\qquad
B=B_1+B_2+B_3,
\]
where the monomials $A_1,A_2,A_3\in\mathcal{M}$ are distinct and the monomials $B_1,B_2,B_3\in\mathcal{M}$ are distinct. The \emph{Bivariate Bicycle code} with parameters $(l,m,A,B)$ is the CSS code whose Tanner graphs are $\mathcal{T}_X$ and $\mathcal{T}_Z$ defined in \cref{BB_codes:Tanner_graph}.
\end{definition}
\section{Error Bound in the Logical Error Rate Estimate}\label{error_bound}
Our method for the logical error rate estimation \ref{eq:LER_estimate} requires bounding the higher weight contribution term
\[
    r = \sum_{w=\bar{w}}^n N_w p^w (1-p)^{n-w},
\]
where $\bar{w} = 6$.\\
To bound $N_w$ considering the worst case estimate $N_w \leq \binom{n}{w}$, and then apply the inequality $\binom{a}{b}\leq \left(\frac{a e}{b}\right)^b$ to obtain
\[
    N_w \leq \binom{n}{w} \leq \left(\frac{n e}{w}\right)^w.
\]
We then manipulate the expression
\begin{multline*}
    r \leq \sum_{w=\bar{w}}^n \left(\frac{n e}{w}\right)^w p^w (1-p) ^{n-w}
    \leq \sum_{w=\bar{w}}^n \left(\frac{n e}{w}\right)^w p^w \leq \\
    \leq \sum_{w=\bar{w}}^n \left(\frac{n e}{\bar{w}}\right)^w p^w
    = \sum_{w=\bar{w}}^n \beta ^ w
\end{multline*}
with $\beta = \frac{n e p }{\bar{w}}$.\\
By summing the geometric series, we obtain
\[
  \sum_{w=\bar{w}}^n \beta ^ w = \beta^{\bar{w}} \sum_{j=0}^{n - \bar{w}} \beta^j = \beta^{\bar{w}} \frac{1-\beta^{n-\bar{w}+1}}{1-\beta}
\]
and therefore the bound 
\begin{equation*}
    r \leq \beta^{\bar{w}} \frac{1-\beta^{n-\bar{w}+1}}{1-\beta}
\end{equation*}
that we used in our simulations.

\section{Proof of \Cref{lemma:symmetric_BP_evolution}}
For convenience, we rewrite the statement of the lemma.

\begin{lemmastar}\label{lemma_proof:symmetric_BP_evolution}
    Suppose $\sigma$ is a symmetry for syndrome $s$. Then, at every iteration of (parallel) BP, the hard decision $\tilde{x}$ satisfies $\sigma(\tilde{x}) = \tilde{x}$.
\end{lemmastar}
\begin{proof}
     Denote with $\mathcal{T} = (V,C,E)$ the Tanner graph which BP is applied to.\\
     The proof is based on the symmetry induced by the automorphism $\sigma$. More precisely, can show inductively on the BP iteration counter $t$ that $m^t_{u\to w} = m^t_{\sigma(u)\to \sigma(w)}$ for every $(u,w)\in E$. 
    
    The base case ($t=0$) follows from the constant initialization of beliefs in the uniform physical error rate. 
     For the inductive step, suppose now that the equality $m_{v\to c}^{t-1} = m_{\sigma(v)\to \sigma(c)}^{t-1}$ holds for iteration $t-1$ and nodes $c\in C$, $v \in V$. For the next round of check-to-bit messages, we have:
    \begin{align*}
            m^t_{c\to v} &= (-1)^{s_c} f^{-1}\left[\prod_{v'\in N(c)\setminus \{v\}} f(m^{t-1}_{v' \to c})\right] \\
    	&=  (-1)^{\sigma(s_c)} f^{-1}\left[\prod_{v'\in N(c)\setminus \{v\}} f(m^{t-1}_{v' \to c}) \right]\\
    	&= (-1)^{\sigma(s_c)} f^{-1}\left[\prod_{v'\in N(c)\setminus \{v\}} f(m^{t-1}_{\sigma(v') \to \sigma(c)}) \right]\\
    	&=  (-1)^{\sigma(s_c)} f^{-1}\left[\prod_{v'\in N(\sigma(c))\setminus \{\sigma(v)\}} f(m^{t-1}_{v' \to c})\right] \\ 
        &=  m^t_{\sigma(c)\to \sigma(v)}
    \end{align*}
    where the second equality holds because $\sigma(s) = s$, and the third one thanks to the inductive hypothesis. As a consequence, for bit-to-check messages, we have:
    \begin{align*}
    	m^t_{v \to c} &= \LLR(p) + \sum_{c'\in N(v)\setminus \{c\}} m^t_{c' \to v} \\
    		&=  \LLR(p) + \sum_{c'\in N(v)\setminus \{c\}} m^t_{\sigma(c’) \to \sigma(v)}  \\
    		&= \LLR(p) + \sum_{c'\in N(\sigma(v))\setminus \{\sigma(c)\}} m^t_{c' \to v} = m^t_{\sigma(v) \to \sigma(c)},
    \end{align*}
    which concludes the inductive proof.

    The symmetry of the exchanged messages translates to a symmetry of the soft-decision, which in turn induces the regularity of BP's hard decision.
\end{proof}

\end{document}